\documentclass[a4paper,11pt]{article}
\usepackage{jheppub}
\usepackage{bbm} 
\usepackage{epstopdf}

\title{\boldmath Aspects of the QCD $\theta$-vacuum}

\author[a]{Thomas Vonk,}
\author[b,c]{Feng-Kun Guo,} 
\author[a,d,e]{Ulf-G. Mei{\ss}ner}

\affiliation[a]{Helmholtz-Institut f\"{u}r Strahlen- und Kernphysik and Bethe Center for Theoretical Physics, Universit\"{a}t Bonn, D-53115 Bonn, Germany}
\affiliation[b]{CAS Key Laboratory of Theoretical Physics,
            Institute of Theoretical Physics, Chinese Academy of Sciences,
            Beijing 100190, China}
\affiliation[c]{School of Physical Sciences, University of Chinese Academy of Sciences,
             Beijing 100049, China}
\affiliation[d]{Institute for
           Advanced Simulation, Institut f\"ur Kernphysik and J\"ulich Center for Hadron Physics,
           Forschungszentrum J\"ulich, D-52425 J\"ulich, Germany}
\affiliation[e]{Tbilisi State University, 0186 Tbilisi, Georgia}

\emailAdd{vonk@hiskp.uni-bonn.de}
\emailAdd{fkguo@itp.ac.cn}
\emailAdd{meissner@hiskp.uni-bonn.de}

\abstract{This paper addresses two aspects concerning the $\theta$-vacuum of Quantum Chromodynamics. First, large-$N_c$ chiral perturbation theory is used to calculate the first two non-trivial cumulants of the distribution of the winding number, i.\,e. the topological susceptibility, $\chi_\mathrm{top}$, and the fourth cumulant, $c_4$, up to next-to-leading order. Their large-$N_c$ scaling is discussed, and  compared to lattice results. It is found that $\chi_\mathrm{top}=\mathcal{O}(N_c^0)$, as known before, and $c_4=\mathcal{O}(N_c^{-3})$, correcting the assumption of $\mathcal{O}(N_c^{-2})$ in the literature. Second, we discuss the properties of QCD at $\theta\sim\pi$ using chiral perturbation theory for the case of $2+1$ light flavors, i.\,e. by taking the strange quark mass heavier than the degenerate up and down quark masses. It is shown that --- in accordance with previous findings for $N_f=2$ and $N_f=3$ mass-degenerate flavors --- in the region $\theta\sim\pi$ two vacuum states coexist, which become degenerate at $\theta=\pi$. The wall tension of the energy barrier between these degenerate vacua is determined as well as the decay rate of a false vacuum.}


\begin{document}
\maketitle
\flushbottom

\section{Introduction}
This paper is a compilation of two different, but related studies on the $\theta$-vacuum of Quantum Chromodynamics (QCD) within the framework of chiral perturbation theory (CHPT). As a consequence of the QCD $\theta$-vacuum, which itself is a result of the  instanton solution~\cite{BPST} (for a review on instantons, see~\cite{schaefershuryak}),  the $\theta$-term appears in the Lagrangian
\[
\mathcal{L}^\theta_\mathrm{QCD} = - \theta w(x)\, ,
\]
where
\[
w(x)=\frac{g^2}{16\pi^2} \operatorname{Tr}\left[G_{\mu\nu}\tilde{G}^{\mu\nu}\right]
\]
is the winding number density such that $\int \mathrm{d}^4x\,w(x) = \nu$ is the winding number of the respective field configuration. $G_{\mu\nu}$ is the QCD field strength tensor defined as usual, and $\tilde{G}_{\mu\nu}=\frac{1}{2}\epsilon_{\mu\nu\rho\sigma}G^{\mu\nu}$ its dual.

The inclusion of this inconspicuous term as well as the instanton solution itself have, however, severe consequences for QCD:
\begin{enumerate}
\item The vacuum structure of QCD is much more complex than originally thought and depends on the distribution of the winding number~\cite{callandashengross1,callandashengross2,leutwylersmilga}.
\item The $\theta$-term and the related chiral phase of the quark masses, which are conventionally collected in the effective vacuum angle $\bar{\theta}=\theta+\operatorname{Arg}\det\mathcal{M}$, where $\mathcal{M}$ is the quark mass matrix, violate $CP$-symmetry, unless $\bar \theta$ would turn out to be exactly zero (or if one quark mass would be zero so that the angle can be rotated away). At $\bar \theta=\pi$, however, $CP$-symmetry would be recovered, but due to the appearance of degenerate vacuum states, $CP$-symmetry can be spontaneously broken \cite{witten1980,smilga}.
\item Experimentally such a $CP$-violation has not been observed. In fact, the measurements of the neutron electric dipole moment~\cite{Baker:2006ts,Afach:2015sja}, which is a $CP$-odd quantity, give a stringent upper bound for $\bar{\theta}$ \cite{baluni,GuoEtAlNEDM,Dragos:2019oxn}
\[
|\bar{\theta}| \lesssim 10^{-11}\, .
\]
The question why $\bar{\theta}$ is such a small quantity demands  further explanation, since this parameter is not determined by the standard model and would naively be expected to be $\mathcal{O}(1)$. This issue is known by the name of ``strong $CP$-problem'' (for reviews on the axion solution to the ``strong $CP$-problem'', see e.\,g. Refs.~\cite{kimaxions,kimaxions2}).
\end{enumerate}
The first study in section~\ref{ch:vacstruct} is related to the first point,  and focuses on the vacuum structure of QCD, which can be studied by means of the cumulants of the topological distribution of the winding number. In particular, we will investigate  how the related quantities scale in the limit of an arbitrarily large numbers of colors $N_c$ \cite{thooftlargenc} (``large-$N_c$ limit'' or ``'t Hooft limit''; see also the reviews \cite{vicari,lucini}). 
This will be done by considering CHPT for the large-$N_c$ limit up to the next-to-leading order, which hence fills a gap in the recent literature (only leading order results are known so far~\cite{leutwylersmilga,lucianomeggiolaro}). The results will then be compared to results obtained in lattice simulations.

In section~\ref{ch:thetasimpi}, the remarkable properties of the theory at $\theta\sim\pi$,\footnote{Here and in the following, the quark masses are taken to be real and we use $\theta$ to represent $\bar \theta$.} which have been briefly mentioned above, will be investigated. This has been done already in the past for the case of two and three mass-degenerate flavors by Smilga~\cite{smilga} and for the case of one and zero light flavors in Ref.~\cite{Tytgat:1999yx}. Here, we will complement these previous findings by studying the case of $2+1$ flavors, i.\,e. by taking the effects of a much heavier strange quark into account. It will be shown that qualitatively (up to a slightly different prefactor due to approximations made during the calculation) Smilga's 3-flavor results are compatible with the results for $2+1$ flavors.

Before embarking upon these studies, note that the present paper on the $\theta$-vacuum deals with fundamental properties of the \emph{low-energy regime} of QCD, which is not amenable by standard perturbation theoretical methods based on an expansion of couplings. This domain is characterized by nonperturbative interactions and confinement meaning that the degrees of freedom are hadrons rather than the fundamental degrees of freedom of the high-energy regime, the (asymptotically free) quarks and gluons. Because of the symmetry patterns of QCD,  a suitable \emph{effective field theory} can be formulated that adequately describes the low-energy sector of QCD and that builds the framework for the succeeding studies: chiral perturbation theory based on the spontaneous breaking of chiral symmetry as developed in Refs.~\cite{gasserleutwyler1,gasserleutwyler2} (for introductions, see, e.\,g. \cite{xptbernardmeissner,Scherer:2012xha}).

\section{Vacuum structure in the large-\texorpdfstring{$N_c$}{Nc} limit}\label{ch:vacstruct}

This section deals with the properties of the distribution of the winding number at $\theta=0$ in QCD. In section~\ref{ch:systematic} the topological susceptibility $\chi_\mathrm{top}$ and the higher order cumulants $c_n$ are introduced, which are quantities characteristic for the topological vacuum structure of QCD. These quantities have been calculated in CHPT up to the leading and next-to-leading orders considering different scenarios (e.\,g., $N_f=2,3$ with different quark masses, where $N_f$ is the number of light quark flavors, or arbitrary $N_f$ with degenerate quark masses; cf. e.\,g. Refs.~\cite{lucianomeggiolaro,maochiu,bernard2012b,bernard2012a,guomeissner}), and have been measured on the lattice (cf. e.\,g. Refs.~\cite{vicari,bonati,Dimopoulos:2018xkm,Ce:2015qha,Ce:2016awn}).

However, as will be shown in section~\ref{ch:largenc}, QCD undergoes certain modifications if one considers the number of colors arbitrarily large instead of $N_c=3$. The topological susceptibility and the fourth cumulant in this large-$N_c$ limit using the leading order CHPT have been already calculated in Refs.~\cite{leutwylersmilga,lucianomeggiolaro} for arbitrary $N_f$ (degenerate quark masses), but an explicit calculation of the respective quantities up to the next-to-leading order is still lacking. The objective of this study hence is to fill this gap and to work out the actual large-$N_c$ behavior of $\chi_\mathrm{top}$ and $c_4$ up to the next-to-leading order (section~\ref{ch:XtopinlargeNc}), which has been claimed to be  $\chi_\mathrm{top}=\mathcal{O}(1)+\mathcal{O}(N_c^{-2})$ and $c_4=\mathcal{O}(N_c^{-2})$~\cite{vicari,bonati}. These claims, however, will be reexamined in section~\ref{ch:largencresults}.

\subsection{Systematic investigation of the vacuum structure}\label{ch:systematic}

A very comprehensive work on the role of the $\theta$-vacuum and the associated winding number in QCD has been provided by Leutwyler and Smilga~\cite{leutwylersmilga}. Some of their numerous insights form the basis of the calculations performed in section~\ref{ch:XtopinlargeNc}, so let us recall those aspects that are relevant for what follows.
Consider the Euclidean action of QCD including the $\theta$-term over a space-time volume $V$
\begin{align*}
S_\mathrm{E} & = \int_V \mathrm{d}^4x\left(\frac{1}{2}\operatorname{Tr}\left[G_{\mu\nu} G_{\mu\nu} \right] -i\theta w(x) +\bar{q}\left(-i\gamma_\mu \mathcal{D}_\mu + \mathcal{M}\right)q\right)\\
& = S_G - i\theta\nu + \int_V \mathrm{d}^4x\, \bar{q}\left(-i\gamma_\mu \mathcal{D}_\mu + \mathcal{M}\right)q \, ,
\end{align*}
where $q$ represents the quark fields, $\mathcal{D}_\mu$ is the QCD gauge covariant derivative, $\mathcal M$ the quark mass matrix, and
$$S_G = \int_V \mathrm{d}^4x\, \frac{1}{2}\operatorname{Tr}\left[G_{\mu\nu} G_{\mu\nu} \right]$$
is the gluon action.

The partition function involving a path integral over gauge-field configurations characterized by the winding number $\nu$ will hence involve a sum over $\nu$ and will depend explicitly on $\theta$:
\begin{align}
Z (\theta) & = \int [\mathrm{D}A_\mu ]\,\exp\left(-S_G + i\theta\nu\right)\operatorname{det}\left(-i\gamma_\mu \mathcal{D}_\mu + \mathcal{M}\right)\nonumber\\
& \equiv \sum_{\nu=-\infty}^{+\infty} e^{i\theta\nu} Z_\nu\ ,\label{eq:partitionfct} 
\end{align}
where 
\[
Z_\nu := \int [\mathrm{D}A_\mu ]_\nu \exp\left(-S_G\right)\operatorname{det}\left(-i\gamma_\mu \mathcal{D}_\mu + \mathcal{M}\right)
\]
is the partition function with given winding number $\nu$, and the functional determinant
\[
\operatorname{det}\left(-i\gamma_\mu \mathcal{D}_\mu + \mathcal{M}\right) = \int [\mathrm{D}q][\mathrm{D}\bar{q}]\,\exp\left(\int_V \mathrm{d}^4x\, \bar{q}\left(-i\gamma_\mu \mathcal{D}_\mu + \mathcal{M}\right)q\right)
\]
has been inserted. With that one  arrives at a probabilistic interpretation
\[
p_\nu=\frac{Z_\nu}{Z(\theta=0)} 
\]
with $p_\nu$ describing the probability to find a field configuration with winding number $\nu$~\cite{leutwylersmilga}. 

Considering the $n^{\rm th}$ derivative of $Z(\theta)$ at $\theta=0$, one finds
\begin{align*}
\langle\nu^n\rangle_{\theta=0} & = \frac{1}{i^n Z(\theta=0)} \left[\frac{\partial^n Z}{\partial \theta^n}\right]_{\theta=0}\,,
\end{align*}
meaning that $Z(\theta)$ can be regarded as the moment-generating function of the distribution of the winding number $\nu$ (in the following paragraphs we will drop the subscript $\theta=0$ of $\langle \nu^n \rangle_{\theta=0}$). Since the mean $\langle \nu \rangle=0$, the $n^{\rm th}$ moment is also the $n^{\rm th}$  central moment, and because the distribution of $\nu$ is symmetric, one can conclude $\langle \nu^{2n+1} \rangle = 0$, $n\in \mathbb{N}_0$. Note that in particular the mean square per unit volume 
\begin{equation*}
\chi_\mathrm{top} := \frac{\langle \nu^2\rangle}{V}
\end{equation*} 
is the topological susceptibility~\cite{crewther1977}. Using the explicit form of the partition function $Z$, the topological susceptibility of pure gluodynamics (GD) can be expressed as
\[
\tau = \frac{\langle \nu^2 \rangle_\mathrm{GD}}{V} = \int \mathrm{d}^4x \langle 0| T w(x) w(0) | 0 \rangle_\mathrm{GD}\, ,
\]
where $T$ denotes the time-ordering operator, which in Minkowski space reads
\begin{equation}\label{eq:tauinminkowski}
\tau = -i \int \mathrm{d}^4x \langle 0| T w(x) w(0) | 0 \rangle_\mathrm{GD}\ .
\end{equation}
The case we are considering here is $V\Sigma m \gg 1$, with $\Sigma$ the absolute value of the quark condensate in the chiral limit (see also below) and $m$ the isospin symmetric up and down quark mass.  In this particular case, the partition function is dominated by the ground state energy $e_\mathrm{vac}(\theta)$, \cite{leutwylersmilga,guomeissner}
\begin{equation*}
Z(\theta) = e^{-Ve_\mathrm{vac}(\theta)}\,,
\end{equation*}
or equivalently
\[
e_\mathrm{vac}(\theta)=-\frac{1}{V}\ln Z(\theta)\, .
\]
Since $Z(\theta)$ is the moment-generating function, and since $e_\mathrm{vac}(\theta)$ is basically its natural logarithm (up to the factor $-1/V$), the latter can be regarded as the cumulant-generating function
\[
c_n = \left[\frac{\partial^n e_\mathrm{vac}}{\partial\theta^n}\right]_{\theta=0}\, .
\]
Note that the second cumulant
\[
c_2 = -\frac{1}{VZ(\theta=0)}\left[\frac{\partial^2 Z}{\partial \theta^2}\right]_{\theta=0}  \equiv \chi_\mathrm{top}
\]
is nothing but the topological susceptibility. The properties of the distribution of the winding number imply that the vacuum energy density is an even function in $\theta$ with the cumulants being the respective coefficients:
\begin{align}
e_\mathrm{vac}(\theta) & = e_\mathrm{vac}(0) + \sum_{n=1}^\infty \frac{c_{2n}}{(2n)!}\theta^{2n}\nonumber\\
& = e_\mathrm{vac}(0) + \frac{\chi_\mathrm{top}}{2!}\theta^2 + \frac{c_4}{4!}\theta^4 + \dots \label{eq:evacexpansion}
\end{align}
Note that, if all $c_n$'s vanish for $n\geq 4$, the distribution of the winding number will be purely Gaussian. For $N_c=3$ this is not the case (there $c_4\ne 0$~\cite{guomeissner}), but as will be shown below, $c_4$ vanishes rapidly in the large-$N_c$ limit (as do all higher cumulants), while $\chi_\mathrm{top}$ remains finite, so that in the large-$N_c$ limit, the distribution of the winding number is indeed Gaussian.

\subsection{The large-\texorpdfstring{$N_c$}{Nc} effective Lagrangian}\label{ch:largenc}

The large-$N_c$ limit has been introduced by 't Hooft~\cite{thooftlargenc}. This theoretical model is  worth a thorough study because the theory undergoes certain simplifications: 't Hooft recognized that if $N_c\to\infty$ and $g^2 N_c=\lambda$ is held fixed, where $g$ is the strong coupling constant, the amplitude $\mathcal{A}$ of any Feynman diagram under consideration comes with a weight factor
\begin{equation*}
\mathcal{A} \sim \lambda^{P-V} N_c^{2-2H-L}\ ,
\end{equation*}
where $P$ is the number of internal propagator lines, $V$ the number of vertices, $H$ the number of ``holes'', and $L$ the number of quark loops. Consequently, the total amplitude of any process under consideration in the large-$N_c$ limit will be dominated by those diagrams that have no ``holes'' $H=0$, i.\,e. planar diagrams, and by those diagrams that come with the minimal number of quark loops $L$.

Furthermore, the theory achieves a higher degree of symmetry due to the fact that the singlet axial current, which is anomalous in the usual QCD, is conserved in the large-$N_c$ limit,
\begin{equation*}
\partial_\mu J_A^\mu = \frac{N_f g^2}{8 \pi^2} \operatorname{Tr}\left[ G_{\mu\nu}\tilde{G}^{\mu\nu}\right] = \frac{N_f \lambda}{8 \pi^2 N_c} \operatorname{Tr}\left[ G_{\mu\nu}\tilde{G}^{\mu\nu}\right] \xrightarrow{N_c \to \infty} 0\, .
\end{equation*}
As a consequence, the large-$N_c$ QCD with its approximate U(3) flavor symmetry comes with a ninth pseudo-Nambu--Goldstone boson, the $\eta^\prime$, associated with the spontaneous symmetry breaking of the axial U(1)$_A$ symmetry.

The leading order effective Lagrangian in the large-$N_c$ limit has been derived independently by Di Vecchia and Veneziano,  Rosenzweig et al., and Witten \cite{witten1980,vecchiaveneziano1980,rosenzweig1980} using slightly different strategies (see also~\cite{kaiserleutwyler,Kawarabayashi:1980dp,Ohta:1981ai}). Consider the case $N_f=3$ with massless quarks, such that the dynamics of the low-energy regime of the theory in the large-$N_c$ limit will be governed by the nine Nambu--Goldstone bosons associated with the symmetry breakdown from U(3)$_L\times$U(3)$_R$ to U(3)$_V$. These nine pseudoscalar fields $\phi_k (x)$, $k=0,\dots ,8$, may  be collected in the matrix $U(x)\in$~U(3) such that additionally the $\eta^\prime = \phi_0(x)\sim \Psi(x)$ shows up in the phase of the determinant of $U(x)$, i.\,e.
\begin{equation}\label{eq:fieldmatrixinu3}
U(x)=\exp\left(i\frac{\lambda_k \phi_k}{F} \right),
\end{equation}
and
\begin{equation}\label{eq:etainphase}
\operatorname{det}U=e^{i\Psi}\, ,
\end{equation}
where $F$ is the pion decay constant in the chiral limit, $\lambda_0=\sqrt{\frac{2}{3}}\mathbbm{1}$, and $\lambda_{k=1,\dots ,8}$ are the Gell-Mann matrices. From eq.~\eqref{eq:etainphase} one can easily derive
\begin{equation}\label{eq:psiphi0}
\Psi = -i\ln\operatorname{det}U = \frac{\sqrt{6}}{F}\phi_0\ .
\end{equation}
Under U(3)$_L\times$U(3)$_R$, the matrix $U(x)$ transforms as
\[
U\to U^\prime = V_R U V_L^\dagger\,,
\]
and because of eq.~\eqref{eq:etainphase},
\begin{equation}\label{eq:psitrafo}
\Psi \to \Psi^\prime = \Psi -i\ln\operatorname{det}V_R+i\ln\operatorname{det}V_L\, .
\end{equation}
As in the usual SU(3) CHPT, one may construct an effective Lagrangian that is compatible with the symmetries under consideration. The leading order Lagrangian will of course come with the same terms as in the SU(3) case, where now $U(x)\in$~U(3). Denote it by $\mathcal{L}_0$.
Moreover, one now needs to consider also the $\theta$-term
\[
\mathcal{L}_\theta = -\theta w(x)
\]
and a term that takes into account the axial anomaly. This term, however, should be invariant under SU(3)$_L\times$SU(3)$_R$, but should be noninvariant under U(3)$_L\times$U(3)$_R$ such that
\begin{equation}\label{eq:Lanomtrafo}
\mathcal{L}_\mathrm{anom.} \to \mathcal{L}_\mathrm{anom.}+N_f \phi w(x)
\end{equation}
as in QCD with $\phi$ the U(1)$_A$ rotation angle. Certainly, a term $\sim \Psi$ is invariant under SU(3)$_L\times$SU(3)$_R$, because according to eq.~\eqref{eq:psitrafo} $\Psi\to\Psi^\prime=\Psi$ if $V_{R/L}\in$ SU(3)$_{R/L}$, since then $\operatorname{det}V_{R/L}=1$. The term that fulfills both conditions is
\[
\mathcal{L}_\mathrm{anom.}= - \Psi w(x) \, .
\]
Another piece that is invariant under U(3)$_L\times$U(3)$_R$ and parity is~\cite{vecchiaveneziano1980}
\[
\mathcal{L}_{w^2} = \frac{1}{2\tau} w^2(x)\, ,
\]
where we have set the coefficient such that in pure gluodynamics at $\theta=0$ eq.~\eqref{eq:tauinminkowski} is reproduced.
Collecting all terms $\mathcal{L}_\mathrm{eff}=\mathcal{L}_0+\mathcal{L}_\theta+\mathcal{L}_\mathrm{anom.}+\mathcal{L}_{w^2}$ yields
\[
\mathcal{L}_\mathrm{eff}=\mathcal{L}_0-w(x)\left(\theta + \Psi\right) +\frac{1}{2\tau}w^2(x)\, .
\]
Using the the classical equation of motion for $w(x)$,
\[
w(x) = \tau \left( \theta + \Psi\right)\, ,
\]
one finally finds
\begin{equation}\label{eq:LefflargeNcLO}
\mathcal{L}_\mathrm{eff} = \frac{F^2}{4}\operatorname{Tr}\left[(\mathrm{D}_\mu U)^\dagger \mathrm{D}^\mu U\right]+\frac{F^2}{2}\operatorname{Re}\operatorname{Tr}\left[\chi^\dagger U\right] - \frac{\tau}{2}\left( \theta + \Psi\right)^2\, ,
\end{equation}
where the first two terms are $\mathcal{L}_0$.
It is crucial to note that the large-$N_c$ counting rules imply that \cite{leutwylersmilga}
\begin{equation}\label{eq:largeNccoefficiets}
F=\mathcal{O}(\sqrt{N_c}),\quad \Sigma = \mathcal{O}(N_c), \quad \tau = \mathcal{O}(N_c^0)\ ,
\end{equation}
where $\Sigma=|\langle 0|\bar{q}q|0\rangle|$ is the absolute value of the quark condensate in the chiral limit (implicitly present in eq.~\eqref{eq:LefflargeNcLO} via $\chi = {2\Sigma}\mathcal{M}/{F^2}$).

Expanding the exponential function in $U$ one finds that it contains vertices of the type $F^{2-n}(\partial\phi)(\partial\phi)\phi^{n-2}$ describing interactions among $n$ pseudoscalar mesons ($n>2$). It thus follows that such interactions proportional to $F^{2-n}=\mathcal{O}(N_c^{1-n/2})$ are suppressed in the large-$N_c$ limit (the more particles participate the stronger this suppression is), meaning that these mesons become free particles if $N_c\to\infty$~\cite{kaiserleutwyler}.

From the effective Lagrangian given above and considering degenerate light quark masses, one may derive the leading order squared mass of the $\eta^\prime$: 
\begin{equation}\label{eq:etaprimemass}
M_{\eta^\prime}^2 = \frac{2\Sigma m}{F^2} + \frac{6\tau}{F^2}\ , 
\end{equation}
which is the  Witten--Veneziano formula~\cite{veneziano}. This formula clearly shows that the $\eta^\prime$ is not massless even in the chiral limit $m=0$. In the joint limit  $m=0$ and $N_c\to\infty$, the ninth Nambu--Goldstone boson finally also becomes massless, because the second term in eq.~\eqref{eq:etaprimemass} scales like $\mathcal{O}(N_c^{-1})$, which follows immediately from eq.~\eqref{eq:largeNccoefficiets}. Calculating the $\eta^\prime$-mass in the $2+1$ flavor CHPT with $m_u=m_d=m < m_s$, one finds
\[
M_{\eta^\prime}^2 = \frac{2\Sigma}{3F^2}(2m+m_s)+ \frac{6\tau}{F^2}\, ,
\]
which gives a more refined version of the Witten--Veneziano formula. Using $M_\eta^2={2\Sigma}(m+2m_s)/{3F^2}$ and $M_K^2={\Sigma}(m+m_s)/{F^2}$, one gets
\begin{equation}\label{eq:wittenvenezianofine}
M_{\eta^\prime}^2+M_\eta^2-2M_K^2=\frac{6\tau}{F^2}\, .
\end{equation}
Clearly this large-$N_c$ behavior of $M_{\eta^\prime}^2$ is a consequence of the suppression of the axial anomaly, which, as has been shown above, is also  of $\mathcal{O}(N_c^{-1})$.

Of course, also higher order large-$N_c$ effective Lagrangians can be constructed. In order to construct such higher order effective Lagrangians one may, however, ask how to include the additional low-energy scale appearing in the large-$N_c$ limit: the mass of the $\eta^\prime$. Kaiser and Leutwyler~\cite{kaiserleutwyler} argued that a coherent analysis comprising this is to treat $M_{\eta^\prime}^2 \sim 1/N_c$ and $p^2$ on equal footing, since both appear in the $\eta^\prime$ propagator, leading to their $\delta$-expansion scheme in which $p^2$, the quark mass $m$, and $1/N_c$ represent all quantities of the same order $\mathcal{O}(\delta)$. However, one may as well extend the common $p$-expansion of SU(3)$_L\times $SU(3)$_R$ to U(3)$_L\times $U(3)$_R$ \cite{HerreraSiklody:1996pm}
and add possible terms $\sim {\bar{\Psi}}^n=(\theta+\Psi)^n$ and $\sim \mathrm{D}_\mu\theta$ which are still consistent with the desired symmetries.
Consequently, the NLO Lagrangian will come with some terms that have the same form as in the  SU(3)$_L\times $SU(3)$_R$ case, meaning that they come with low-energy constants (LECs) $L_1,\dots ,L_{10}$, and high-energy constants (HECs) $H_1$ and $H_2$ for the contact terms,\footnote{The coefficients $L_i$ and $H_i$, as well as $F$ and $B$ of the extended U(3) theory have to be distinguished from the corresponding ones appearing in the SU(3) case, but they can be matched, as has been shown in Ref.~\cite{kaiserleutwyler}.} some new terms of $D_\mu\theta$ with LECs $L_{11},\dots, L_{17}$ and HECs $H_3, \dots, H_6$, and some new terms of $\bar{\Psi}$ with coefficients $\Lambda_1$ and $\Lambda_2$. Here we  refrain from giving the whole next-to-leading order Lagrangian, instead the relevant pieces will be presented below in the respective sections, see sec.\,\ref{ch:deltaexpansion} and sec.~\ref{ch:NLOpexpansion}.

For determining the large-$N_c$ scaling of the LECs and HECs of the respective terms in the Lagrangian, Kaiser and Leutwyler~\cite{kaiserleutwyler} found a very simple power counting rule, which is based, of course, on the corresponding large-$N_c$ power counting scheme of QCD. One just counts the number of flavor traces $n_{\mathrm{Tr}}$, the number of $\bar{\Psi}$-factors $n_{\bar{\Psi}}$, and the number $n_{\mathrm{D}_\mu\theta}$ of factors with $D_\mu\theta$ appearing in a given term of the effective Lagrangian, then the associated coefficient would be
\begin{equation}\label{eq:largeNcCoefficients}
C = \mathcal{O}(N_c^{2-k})\ ,\qquad k=n_\mathrm{Tr}+n_{\bar{\Psi}}+n_{\mathrm{D}_\mu\theta}\, ,
\end{equation}
where $C$ may be any of the above mentioned coefficients $F,B,L_i,H_i,\Lambda_i$ (or their combinations).

Before embarking on the explicit determination of the large-$N_c$ behavior of the topological susceptibility and the fourth cumulant, one may ask whether there may be any way to predict the large-$N_c$ behavior of these quantities. For that, note that in the large-$N_c$ limit the actual relevant ordering parameter in the expression for $e_\mathrm{vac}$ is not $\theta$, but $\vartheta=\theta/N_c$ \cite{witten1980}:
\begin{equation}\label{eq:evacexpansionlargeNc}
e_\mathrm{vac}(\vartheta)-e_\mathrm{vac}(0) = N_c^2\left(\frac{\chi_\mathrm{top}}{2}\vartheta^2 + \frac{c_4}{4!}N_c^2\vartheta^4+\dots\right)\left(1+\mathcal{O}(N_c^{-1})\right) ,
\end{equation}
cf. eq.\,\eqref{eq:evacexpansion} above. This  implies that $\chi_\mathrm{top}$ is at most of order $\mathcal{O}(1)+\mathcal{O}(N_c^{-1})$, whereas $c_4$ should be {\em at most} of order $\mathcal{O}(N_c^{-2}) +\mathcal{O}(N_c^{-3})$. 
In the next sections it is shown that the leading order and next-to-leading order results for these quantities are in accordance with these expected maximal large-$N_c$ scalings.

\subsection{The topological susceptibility and fourth cumulant in the large-\texorpdfstring{$N_c$}{Nc} limit}\label{ch:XtopinlargeNc}

In this section, we first reproduce the leading order results for $\chi_\mathrm{top}$ and $c_4$ that have been already derived before~\cite{leutwylersmilga,lucianomeggiolaro}. These leading order results will then be complemented by the next-to-leading order calculations of both quantities, considering both the $\delta$-expansion of Kaiser and Leutwyler~\cite{kaiserleutwyler} and the standard $p$-expansion of  chiral perturbation theory. It is shown that both approaches lead to the same large-$N_c$ scaling of the respective quantities.

Note that in what follows, we assume  the U($N_f$) symmetric case, i.\,e. we consider an arbitrary number of flavors with equal masses $m > 0$, and a mass matrix that is real, positive, and diagonal $\mathcal{M}=m \mathbbm{1}_{N_f \times N_f}$.

\subsubsection{Leading order}

The potential of the leading order Lagrangian \eqref{eq:LefflargeNcLO} is given by
\begin{equation*}
V(U,\theta) = - \Sigma  \operatorname{Re} \operatorname{Tr}\left[\mathcal{M}U^\dagger\right] + \frac{\tau}{2}\left(\Psi +\theta\right)^2\, ,
\end{equation*}
The vacuum energy density is then found by minimizing this potential with respect to $U$. However, with the diagonal mass matrix in the U($N_f$) symmetric case, the minimum will occur when $U$ is a multiple of the unit matrix, $U=\exp\left(i{\Psi}/{N_f}\right)\mathbbm{1}$:
\begin{equation}\label{evaclo}
e_\mathrm{vac}^\mathrm{LO} = \underset{\Psi}{\mathrm{Min}} \left\{- \Sigma m N_f \cos \frac{\Psi}{N_f} + \frac{\tau}{2} \left(\Psi + \theta\right)^2\right\} .
\end{equation}
For $\theta = 0$, the minimum apparently occurs at $\Psi=0$. In order to find a solution that is analytically manageable one may hence consider the (quite realistic) case $|\theta |\ll 1$, for which $\Psi$ will be very small, too, so one may approximate the cosine in eq.~\eqref{evaclo} up to $\mathcal{O}(\Psi^4)$ (this order is needed to calculate also the fourth cumulant $c_4$):
\begin{equation}\label{evaclo2}
e_\mathrm{vac}^\mathrm{LO} = \underset{\Psi}{\mathrm{Min}} \left\{- \Sigma m N_f \left(1 - \frac{1}{2}\left(\frac{\Psi}{N_f}\right)^2 + \frac{1}{24}\left(\frac{\Psi}{N_f}\right)^4 \right) + \frac{\tau}{2} \left(\Psi + \theta\right)^2 + \mathcal{O}\left(\Psi^6\right)\right\} .
\end{equation}
The minimum appears at
\[
\Psi = -\frac{\tau N_f}{\Sigma m + \tau N_f} \theta + \mathcal{O}\left(\theta^3\right) .
\]
Inserting this into eq.~\eqref{evaclo2} yields
\[
e_\mathrm{vac}^\mathrm{LO} (\theta) = \mathrm{const} + \frac{1}{2}\frac{\Sigma m \tau}{\Sigma m + \tau N_f}\theta^2 - \frac{1}{24}\frac{\Sigma m N_f \tau^4}{(\Sigma m + \tau N_f)^4} \theta^4 + \mathcal{O}\left(\theta^6\right) .
\]
From that the topological susceptibility $\chi_\mathrm{top}^\mathrm{LO}$ and the fourth cumulant $c_{4}^\mathrm{LO}$ are obtained by calculating the respective derivatives:
\[
\chi_\mathrm{top}^\mathrm{LO} = \left[\frac{\partial^2}{\partial \theta^2} e_\mathrm{vac}^\mathrm{LO}(\theta)\right]_{\theta=0} = \frac{\Sigma m \tau}{\Sigma m + \tau N_f}\, ,
\]
which agrees with the result derived already by Leutwyler and Smilga~\cite{leutwylersmilga}. The fourth cumulant is given by
\[
c_{4}^\mathrm{LO} = \left[\frac{\partial^4}{\partial \theta^4} e_\mathrm{vac}^\mathrm{LO}(\theta)\right]_{\theta=0} = - \frac{\Sigma m N_f \tau^4}{(\Sigma m + \tau N_f)^4}\,.
\]
We note that the results reported in eqs.~(4.16) and (4.17) of  \cite{lucianomeggiolaro} are in accordance with the results found here.\footnote{Different notation is used in Ref.~\cite{lucianomeggiolaro}: $\Sigma$ and $\tau$ here correspond to their ${F_\pi B_m}/{2}$ and $A$, respectively.} All variables in these equations are $\mathcal{O}(1)$ with respect to $N_c$ except $\Sigma$ which is $\mathcal{O}(N_c)$. Consequently the final result reads
\begin{equation}\label{eq:loresult}
\begin{split}
\chi_\mathrm{top}^\mathrm{LO} & \xrightarrow{N_c \to \infty} \tau = \mathcal{O}\left(1\right),\\
c_{4}^\mathrm{LO} & \xrightarrow{N_c \to \infty} -\frac{N_f \tau^4}{(\Sigma m)^3} = \mathcal{O}\left(N_{c}^{-3}\right) .
\end{split}
\end{equation}

\subsubsection{Next-to-leading order I: \texorpdfstring{$\delta$}{Delta}-expansion}\label{ch:deltaexpansion}

In Refs.~\cite{Leutwyler:1996np,kaiserleutwyler} a scheme was proposed for constructing $\mathcal{L}_\mathrm{eff}$ in the large-$N_c$ limit by simultaneously expanding in powers of momenta and in powers of $1/N_c$ introducing an ordering parameter $\delta$. As already stated above, terms representing certain powers of momenta $p$ and powers of $1/N_c$, respectively, are counted in this scheme according to
\[
\partial_\mu = \mathcal{O}(\sqrt{\delta})\,,\quad m = \mathcal{O}(\delta)\,, \quad 1/N_c = \mathcal{O}(\delta)\, .
\]
In contrast to usual SU($N_f$) CHPT, where loop graphs are already relevant at the next-to-leading order, loop graphs are relegated to the next-to-next-to-leading order in the large-$N_c$ limit, since graphs containing $L$ loops are of order $p^{2L}$, but are also inversely proportional to powers of $F$, which is of order $\sqrt{N_c}$. As a result, one loop graphs are hence of $\mathcal{O}(\delta^2)$.
The effective Lagrangian up to the next-to-leading order in the $\delta$-expansion is given by \cite{kaiserleutwyler}
\begin{equation*}
\begin{split}
\mathcal{L}_\mathrm{eff}^\delta = & f\left\{\partial_\mu U, \partial_\mu U^\dagger ,\partial_\mu \Psi, \partial_\mu \theta\right\}\\
& + \frac{F^2}{4}\operatorname{Tr}\left[U^\dagger \chi + \chi^\dagger U\right] - \frac{\tau}{2}\left(\Psi +\theta\right)^2 + L_8 \operatorname{Tr}\left[U^\dagger \chi U^\dagger \chi + \chi^\dagger U \chi^\dagger U\right]\\
& -\frac{i}{12}F^2\Lambda_2\left(\Psi + \theta\right)\operatorname{Tr}\left[U^\dagger \chi - \chi^\dagger U\right] + H_2 \operatorname{Tr}\left[\chi\chi^\dagger\right] ,
\end{split}
\end{equation*}
which includes all terms up to $\mathcal{O}(\delta)$ corresponding to terms of $\mathcal{O}(N_c p^4)$, $\mathcal{O}(p^2)$, and $\mathcal{O}(N^{-1}_c)$. The large-$N_c$ scalings of the coefficients follow from eq.~\eqref{eq:largeNcCoefficients}: $\tau=\mathcal{O}(1)$, $\Lambda_2=\mathcal{O}(N^{-1}_c)$, and $(\Sigma,F^2,L_8,H_2) =\mathcal{O}(N_c)$. Terms that are not of interest here have been collected for the sake of brevity in the function $f\left\{\partial_\mu U, \partial_\mu U^\dagger ,\partial_\mu \Psi, \partial_\mu \theta\right\}$. Inserting again $\chi = {2\Sigma}\mathcal{M}/{F^2}$ yields
\begin{equation*}
\begin{split}
\mathcal{L}_\mathrm{eff}^\delta = & f\left\{\partial_\mu U, \partial_\mu U^\dagger ,\partial_\mu \Psi, \partial_\mu \theta\right\}\\
& + \Sigma\operatorname{Re}\operatorname{Tr}\left[U^\dagger \mathcal{M}\right] - \frac{\tau}{2}\left(\Psi +\theta\right)^2 + 8 \frac{\Sigma^2}{F^4} L_8 \operatorname{Re}\operatorname{Tr}\left[U^\dagger \mathcal{M} U^\dagger \mathcal{M}\right]\\
& +\frac{1}{3}\Sigma\Lambda_2\left(\Psi + \theta\right)\operatorname{Im}\operatorname{Tr}\left[U^\dagger \mathcal{M}\right] + \mathrm{const}\, .
\end{split}
\end{equation*}
Using  $\mathcal{M}=m\mathbbm{1}$ and $U=\exp\left(i{\Psi}/{N_f}\right)\mathbbm{1}$ leads to
\begin{equation}
\begin{split}
e_\mathrm{vac}^\delta = \underset{\Psi}{\mathrm{Min}} \biggl\{\biggr. & \mathrm{const} - \Sigma m N_f \cos\left(\frac{\Psi}{N_f}\right) + \frac{\tau}{2} \left(\Psi + \theta\right)^2 \\
& - 8\frac{(\Sigma m)^2 N_f}{F^4}L_8 \cos\left(\frac{2\Psi}{N_f}\right) + \frac{1}{3}\Sigma m N_f \Lambda_2 \left(\Psi + \theta\right)\sin\left(\frac{\Psi}{N_f}\right)\biggl. \biggr\}
\end{split}
\end{equation}
for the vacuum energy density. Following the same argumentation as in the previous section, i.\,e. considering $|\theta|\ll 1$, the sine and the cosines may be expanded up to $\mathcal{O}(\Psi^4)$:
\begin{equation}\label{eq:evacdelta}
\begin{split}
e_\mathrm{vac}^\delta = \underset{\Psi}{\mathrm{Min}} \Biggl\{\Biggr. & - \Sigma m N_f \left(1-\frac{1}{2}\left(\frac{\Psi}{N_f}\right)^2+\frac{1}{24}\left(\frac{\Psi}{N_f}\right)^4\right)
 + \frac{\tau}{2} \left(\Psi + \theta\right)^2 \\
& - 8\frac{(\Sigma m)^2 N_f}{F^4}L_8 \left(1-2\left(\frac{\Psi}{N_f}\right)^2+\frac{2}{3}\left(\frac{\Psi}{N_f}\right)^4\right)  \\
& + \frac{1}{3}\Sigma m N_f \Lambda_2 \left(\Psi + \theta\right)\left(\frac{\Psi}{N_f} - \frac{1}{6}\left(\frac{\Psi}{N_f}\right)^3\right) + \mathcal{O}\left(\Psi^5\right) + \mathrm{const}\Biggl. \Biggr\}\, ,
\end{split}
\end{equation}
which is minimized at
\begin{align}
\Psi & = -N_f \frac{\tau+\frac{1}{3}\Sigma m \Lambda_2}{\Sigma m+\tau N_f + 32 \frac{(\Sigma m)^2}{F^4}L_8+\frac{2}{3}\Sigma m N_f\Lambda_2} \theta + \mathcal{O}\left(\theta^3\right)\nonumber\\
& \equiv -N_f \frac{x}{d} \theta + \mathcal{O}\left(\theta^3\right),
\label{eq:loesungpsidelta}
\end{align}
where we have introduced the abbreviations
\begin{equation}\label{eq:xddelta}
\begin{split}
x & := \tau+\frac{1}{3}\Sigma m \Lambda_2 = \mathcal{O}(1)\,,\\
d & := \Sigma m+\tau N_f + 32 \frac{(\Sigma m)^2}{F^4}L_8+\frac{2}{3}\Sigma m N_f\Lambda_2 = \mathcal{O}(N_c)\, .
\end{split}
\end{equation}
In order to calculate $\chi_\mathrm{top}$, it is reasonable to first consider only those terms in eq.~\eqref{eq:evacdelta} that are quadratic in $\theta$ after the insertion of eq.~\eqref{eq:loesungpsidelta}:
\begin{equation*}
\begin{split}
e_\mathrm{vac}^\delta (\theta) & = \frac{1}{2}\theta^2 \Biggl\{\Biggr. N_f \left(\frac{x}{d}\right)^2\underbrace{\left[\Sigma m + \tau N_f + 32 \frac{(\Sigma m)^2}{F^4} L_8 + \frac{2}{3}\Sigma m \Lambda_2 N_f\right]}_{=d}\\
& \qquad - 2 N_f \frac{x}{d}\underbrace{\left[\tau + \frac{1}{3} \Sigma m \Lambda_2\right]}_{=x} + \tau \Biggl.\Biggr\} + \mathrm{const} + \mathcal{O}\left(\theta^4\right)\\
& = \frac{1}{2}\theta^2\left\{\tau - N_f \frac{x^2}{d} \right\} + \mathrm{const} + \mathcal{O}\left(\theta^4\right) .
\end{split}
\end{equation*}
From that the topological susceptibility follows immediately as
\[
 \chi_\mathrm{top}^\delta = \tau - N_f \frac{x^2}{d}\, .
\]
For the calculation of the fourth cumulant terms quartic in $\theta$ have to be considered:
\begin{equation*}
\begin{split}
e_\mathrm{vac}^\delta (\theta) & = - \frac{\Sigma m N_f}{24}\left(\frac{x}{d}\right)^4\theta^4 - \frac{16(\Sigma m)^2 N_f L_8}{F^4} \left(\frac{x}{d}\right)^4\theta^4 - \frac{1}{18}\Sigma m N_f \Lambda_2 \left(N_f \frac{x}{d} - 1\right) \left(\frac{x}{d}\right)^3\theta^4\\
& \qquad + \mathrm{const} + f(\theta^2) + \mathcal{O}\left(\theta^6\right)\\
& = - \frac{\Sigma m N_f}{24}\left(\frac{x}{d}\right)^4\theta^4\left\{1 + 128\frac{\Sigma m L_8}{F^4} + \frac{4}{3}\Lambda_2\left(\frac{N_f x - d}{x}\right)\right\}\\
& \qquad + \mathrm{const} + f(\theta^2) + \mathcal{O}\left(\theta^6\right),
\end{split}
\end{equation*}
which yields for the fourth cumulant
\[
c_4^\delta = - \frac{\Sigma m N_f x^4}{d^4}\left\{1 + 128\frac{\Sigma m L_8}{F^4} + \frac{4}{3}\Lambda_2\left(\frac{N_f x - d}{x}\right)\right\} .
\]
The behaviors of both the topological susceptibility and the fourth cumulant in the large-$N_c$ limit follow by considering the behaviors of $x$ and $d$ given in eq.~\eqref{eq:xddelta}:
\begin{equation}\label{eq:chitopdeltaresult}
\chi_\mathrm{top}^\delta \xrightarrow{N_c \to \infty} \tau + \mathcal{O}\left(N^{-1}_c\right)\ .
\end{equation}
For the fourth cumulant some further rearrangements lead to a compact expression:
\begin{equation}\label{eq:c4delta}
c_{4}^\delta = -\frac{N_f x^4}{(\Sigma m)^3}\frac{1 + 128\frac{\Sigma m L_8}{F^4} - \frac{4}{3} \Lambda_2 \frac{d}{x} + \frac{4N_f}{3}\Lambda_2 }{\left(1+\frac{\tau N_f}{\Sigma m}+32\frac{\Sigma m L_8}{F^4}+\frac{2 N_f \Lambda_2}{3}\right)^4} \xrightarrow{N_c \to \infty} \mathcal{O}\left(N_{c}^{-3}\right)+ \mathcal{O}\left(N_{c}^{-4}\right).
\end{equation}
The fourth cumulant hence shows the same $N_{c}^{-3}$ suppression as has been already evident in the leading order calculation above (see eq.~\eqref{eq:loresult}), which follows from the fact that all correction terms that show up, i.\,e. terms proportional to $\Sigma m \Lambda_2$ and ${\Sigma m L_8}/{F^4}$ in eq.~\eqref{eq:c4delta}, are of $\mathcal{O}(1)$. The next-to-leading order calculations for both quantities, however, make explicit that the corrections to the leading order term results are of $\mathcal{O}\left(N^{-1}_c\right)$.

\subsubsection{Next-to-leading order II: Full NLO-Lagrangian}\label{ch:NLOpexpansion}

However, it is advisable to check explicitly whether these results can be confirmed by considering the full next-to-leading order Lagrangian from usual CHPT including the terms that represent the effects of the $\eta^\prime$. This effective Lagrangian comes with additional terms proportional to the low-energy constants $L_6$, $L_7$, and $L_{25}$, which in the $\delta$-expansion scheme are relegated to the next-to-next-to-leading order.

The full effective NLO Lagrangian reads (cf. \cite{gasserleutwyler2,kaiserleutwyler}):
\begin{eqnarray}
\mathcal{L}_\mathrm{eff}^\mathrm{NLO} & = & f\left\{\partial_\mu U, \partial_\mu U^\dagger ,\partial_\mu \Psi, \partial_\mu \theta\right\} + \mathrm{const} + \Sigma\operatorname{Re}\operatorname{Tr}\left[U^\dagger \mathcal{M}\right] - \frac{\tau}{2}\left(\Psi +\theta\right)^2 \nonumber\\
&  + & 16 \frac{\Sigma^2}{F^4}L_6\left(\operatorname{Re}\operatorname{Tr}\left[U^\dagger \mathcal{M}\right]\right)^2 - 16 \frac{\Sigma^2}{F^4}L_7\left(\operatorname{Im}\operatorname{Tr}\left[U^\dagger \mathcal{M}\right]\right)^2\nonumber \\ 
& + & 8 \frac{\Sigma^2}{F^4} L_8 \operatorname{Re}\operatorname{Tr}\left[U^\dagger \mathcal{M} U^\dagger \mathcal{M}\right] - 8 \frac{\Sigma^2}{F^4} L_{25}(\Psi + \theta)\operatorname{Im}\operatorname{Tr}\left[U^\dagger \mathcal{M} U^\dagger \mathcal{M}\right]\nonumber\\
&  + & \frac{1}{3}\Sigma\Lambda_2\left(\Psi + \theta\right)\operatorname{Im}\operatorname{Tr}\left[U^\dagger \mathcal{M}\right] .
\end{eqnarray}
The large-$N_c$ scaling for the coefficients of the terms that now show up additionally in comparison to the $\delta$-expansion are: $(L_6,L_7,L_{25}) =\mathcal{O}(1)$. The vacuum energy density is given by
\begin{eqnarray}\label{eq:evacnlo}
e_\mathrm{vac}^\mathrm{NLO} & = &\underset{\Psi}{\mathrm{Min}} \Biggl\{\Biggr. \mathrm{const} - \Sigma m N_f \cos\left(\frac{\Psi}{N_f}\right) + \frac{\tau}{2}\left(\Psi + \theta\right)^2\nonumber \\ 
& &\qquad\qquad -16\frac{(\Sigma m N_f)^2}{F^4}\left[L_6\cos^2\left(\frac{\Psi}{N_f}\right)-L_7\sin^2\left(\frac{\Psi}{N_f}\right)+\frac{L_8}{2N_f}\cos\left(\frac{2\Psi}{N_f}\right)\right]\nonumber \\ 
& & \qquad\qquad - 8 \frac{(\Sigma m)^2 N_f}{F^4} L_{25}\left(\Psi+\theta\right)\sin\left(\frac{2\Psi}{N_f}\right) + \frac{1}{3} \Sigma m N_f \Lambda_2 \left(\Psi+\theta\right)\sin\left(\frac{\Psi}{N_f}\right)\Biggl.\Biggr\},\nonumber 
\end{eqnarray}
which for $\Psi\ll1$ is reduced to
\begin{eqnarray}
e_\mathrm{vac}^\mathrm{NLO} & \stackrel{\Psi\ll 1}{=}& \underset{\Psi}{\mathrm{Min}} \Biggl\{\Biggr. \mathrm{const} - \Sigma m N_f \left[ \frac{1}{2}\left(\frac{\Psi}{N_f}\right)^2 - \frac{1}{24}\left(\frac{\Psi}{N_f}\right)^4\right] + \frac{\tau}{2}\left(\Psi + \theta\right)^2\nonumber\\
& & \qquad\qquad + 16\frac{(\Sigma m N_f)^2}{F^4}\left(L_6 + L_7 + \frac{L_8}{N_f}\right)\left[\left(\frac{\Psi}{N_f}\right)^2 - \frac{1}{3}\left(\frac{\Psi}{N_f}\right)^4\right]\nonumber\\
& & \qquad\qquad - 16 \frac{(\Sigma m)^2N_f}{F^4}L_{25}\left(\Psi + \theta\right)\left[\frac{\Psi}{N_f}-\frac{2}{3}\left(\frac{\Psi}{N_f}\right)^3\right]\nonumber \\
& & \qquad\qquad + \frac{1}{3} \Sigma m N_f \Lambda_2 \left(\Psi+\theta\right)\left[\frac{\Psi}{N_f}-\frac{1}{6}\left(\frac{\Psi}{N_f}\right)^3\right] + \mathcal{O}\left(\Psi^5\right) \Biggl.\Biggr\} .
\end{eqnarray}
The minimization problem is solved by
\begin{align}
\Psi &  = -N_f \frac{\tau-16\frac{(\Sigma m)^2}{F^4}L_{25}+\frac{1}{3}\Sigma m \Lambda_2}{\Sigma m+\tau N_f + 32 \frac{(\Sigma m)^2 N_f}{F^4}\left(L_6+L_7+\frac{L_8}{N_f}-L_{25}\right)+\frac{2}{3}\Sigma m N_f\Lambda_2} \theta+ \mathcal{O}\left(\theta^3\right)\nonumber\\
& \equiv -N_f \frac{X}{D} \theta + \mathcal{O}\left(\theta^3\right) , \label{eq:loesungpsinlo}
\end{align}
where we have once again introduced some abbreviations $X$ and $D$ that are related to the corresponding abbreviations $x$ and $d$ of the $\delta$-expansion calculation above, see eq.~\eqref{eq:xddelta}, by the addition of the term $-16L_{25}{(\Sigma m)^2}/{F^4}=\mathcal{O}(1)$ to $x$ (hence $X$ too is of $\mathcal{O}(1)$) and by replacing $L_8$ in $d$ by $\left(N_f L_6+N_f L_7+L_8-N_f L_{25}\right)$. In the large-$N_c$ limit, however, both $d$ and $D$ are given by the same expression $\Sigma m \left(1+32L_8{\Sigma m}/{F^4}\right)=\mathcal{O}(N_c)$. 

Inserting the solution \eqref{eq:loesungpsinlo} into eq.~\eqref{eq:evacnlo} and considering again only terms $\sim \theta^2$ in order to concentrate first on the calculation of the topological susceptibility results in
\begin{equation*}
e_\mathrm{vac}^\mathrm{NLO} (\theta) = \frac{1}{2}\theta^2\left\{\tau - N_f\frac{X^2}{D}\right\}+ \mathrm{const} + \mathcal{O}\left(\theta^4\right) .
\end{equation*}
This readily gives
\[
 \chi_\mathrm{top}^\mathrm{NLO} = \tau - N_f\frac{X^2}{D}
\]
for the topological susceptibility. Since $X=\mathcal{O}(1)$ and $D=\mathcal{O}(N_c)$ this expression shows the same large-$N_c$ behavior as $X_\mathrm{top}^\delta$, i.\,e.
\begin{equation}\label{eq:chitopnloresult}
 \chi_\mathrm{top}^\mathrm{NLO} \xrightarrow{N_c \to \infty} \tau - \mathcal{O}\left(N^{-1}_c\right) .
\end{equation}
Now we turn to the calculation of the fourth cumulant. For that consider the terms $\sim \theta^4$ in eq.~\eqref{eq:evacnlo} after the insertion of the solution \eqref{eq:loesungpsinlo}:
\begin{equation*}
\begin{split}
e_\mathrm{vac}^\mathrm{NLO} = -\frac{1}{24}\theta^4 \Biggl\{\Biggr.& \Sigma m N_f \left(\frac{X}{D}\right)^4 + 128 \frac{(\Sigma m N_f)^2}{F^4}\left(L_6+L_7+\frac{L_8}{N_f}\right)\left(\frac{X}{D}\right)^4\\
& -256 \frac{(\Sigma m)^2 N_f}{F^4}L_{25}\left(N_f\frac{X}{D}-1\right)\left(\frac{X}{D}\right)^3\\
& +\frac{4}{3}\Sigma m N_f\Lambda_2 \left(N_f\frac{X}{D}-1\right)\left(\frac{X}{D}\right)^3\Biggl.\Biggr\}+\mathrm{const}+f(\theta^2)+\mathcal{O}(\theta^6)\,.
\end{split}
\end{equation*}
This results in the following expression for the fourth cumulant, from which one can derive its large-$N_c$ behavior straightaway:
\begin{equation}\label{eq:c4nloresult}
\begin{split}
c_4^\mathrm{NLO} & = -\Sigma m N_f \left(\frac{X}{D}\right)^4\\
& \quad \Biggl\{ 1+128\frac{\Sigma m}{F^4}\left[N_f L_6+N_f L_7+L_8-2 L_{25}\left(\frac{N_f X-D}{X}\right)\right]+\frac{4}{3}\Lambda_2\left(\frac{N_f X-D}{X}\right)\Biggr\}\\
& \xrightarrow{N_c \to \infty} \mathcal{O}(N_c^{-3}) + \mathcal{O}(N_c^{-4})\,.
\end{split}
\end{equation} 
A comparison with the expression for the fourth cumulant in the $\delta$-expansion scheme, eq.~\eqref{eq:c4delta}, reveals that the results for the large-$N_c$ behavior are basically the same, which is a consequence of the facts that  $x$ and $d$, and, respectively, $X$ and $D$ show the same large-$N_c$ behavior, and that the only additional term in eq.~\eqref{eq:c4nloresult} that persists when sending $N_c$ to infinity, i.\,e. the term $\sim L_{25}$, does not change the overall large-$N_c$ scaling.

\subsection{Comparison with results from lattice simulations}\label{ch:largencresults}

The large-$N_c$ behavior of $\chi_\mathrm{top}=\mathcal{O}(1)+\mathcal{O}(N_c^{-1})$ and $c_4=\mathcal{O}(N_c^{-3})+\mathcal{O}(N_c^{-4})$ is indeed in accordance with the expected scalings that are allowed by eq.~\eqref{eq:evacexpansionlargeNc}. While the former remains finite, the latter is strongly suppressed in the large-$N_c$ limit meaning that the distribution of the winding number becomes purely Gaussian for $N_c\to\infty$.

\begin{figure}[tbh]
\centering
\includegraphics[width=0.75\textwidth]{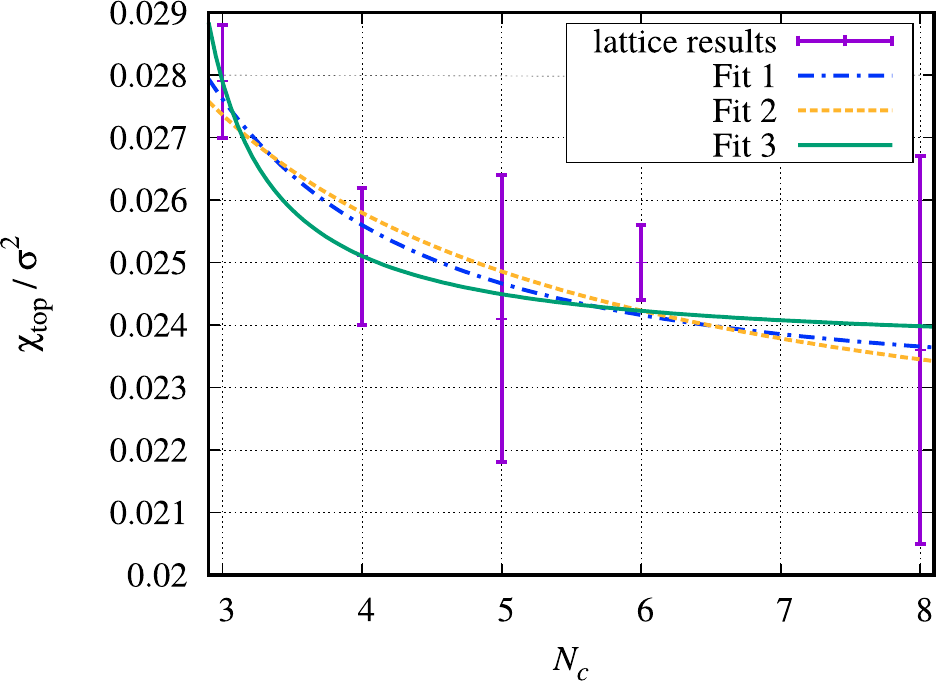}
\caption{Fits to the topological susceptibility measured on the lattice for different values of $N_c$ (lattice results averaged from \cite{vicari,bonati,Lucini:2001ej,DelDebbio:2002xa,Cundy:2002hv,Lucini:2004yh,Ce:2015qha,Ce:2016awn}). Fit~1: $\chi_\mathrm{top}(N_c)=\mathrm{const.}+{a_1}/{N_c^2}$; Fit~2: $\chi_\mathrm{top}(N_c)=\mathrm{const.}+{a_2}/{N_c}$; Fit~3: $\chi_\mathrm{top}(N_c)=\mathrm{a_3}-{(3a_3+a_4)}/{(a_5N_c + 3a_3 +2a_4)}$, where the $a_i$'s are fit parameters.}
\label{fig:chilattice}
\end{figure}

\begin{figure}[tbh]
\centering
\includegraphics[width=0.75\textwidth]{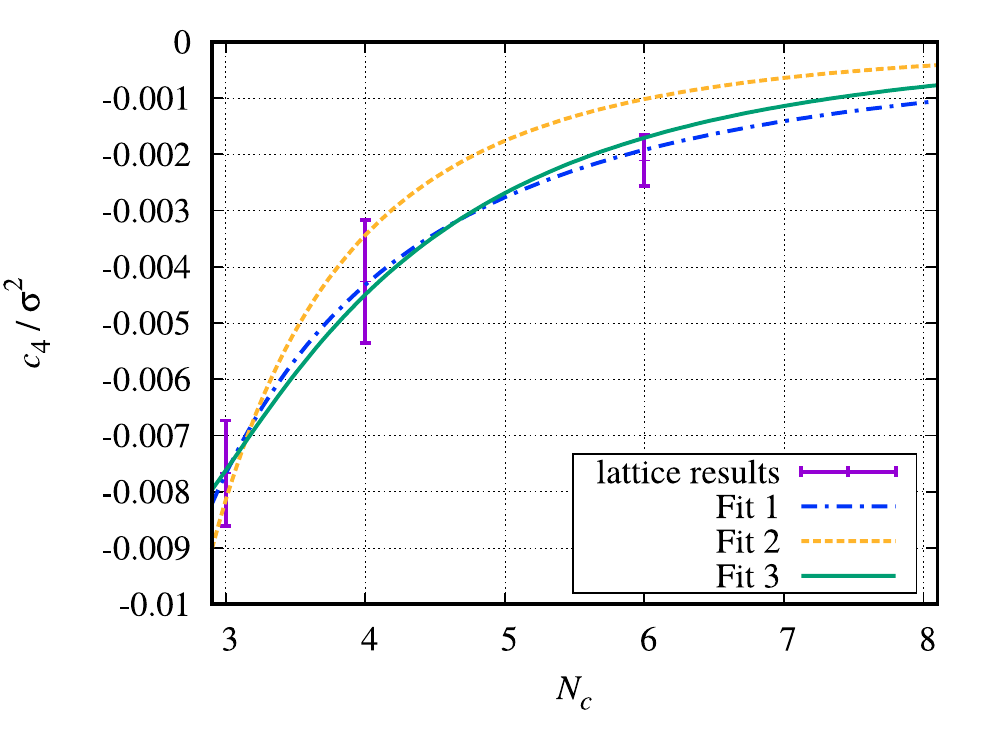}
\caption{Fits to the fourth cumulant measured on the lattice for different values of $N_c$ (lattice results averaged from \cite{bonati,DelDebbio:2002xa,DElia:2003zne,Giusti:2007tu,Ce:2015qha}). Fit~1: $c_4(N_c)={a_6}/{N_c^2}$; Fit~2: $c_4(N_c)={a_7}/{N_c^3}$; Fit~3: $c_4(N_c)=-{(3(a_8+a_9)^4)}/{N_c^3} \cdot {\left[1+{4a_{10}}/{N_c}-4a_9\left(1+{a_{10}}/{N_c}\right)\right]}/{\left(1+{a_{10}}/{N_c}\right)}$, where the $a_i$'s are fit parameters.}
\label{fig:c4lattice}
\end{figure}

The results show, however, a discrepancy to the results in Refs.~\cite{vicari,bonati}. In both papers, the authors based their argumentation on the same large-$N_c$ expression for the vacuum energy density as used here (eq.\,\eqref{eq:evacexpansionlargeNc}, in their notation $b_2=c_4/(12\chi_\mathrm{top})$), but conclude that from this expression it follows that $\chi_\mathrm{top}=\mathcal{O}(1)+\mathcal{O}(N_c^{-2})$ and $c_4=\mathcal{O}(N_c^{-2})+\mathcal{O}(N_c^{-4})$. However, as argued above, eq.~\eqref{eq:evacexpansionlargeNc} allows only to derive an \emph{upper bound} for the scalings of both quantities. Both papers come with lattice calculations for $\chi_\mathrm{top}$ and $b_2=c_4/(12\chi_\mathrm{top})$ for different $N_c$ ranging from $N_c=3$ to $N_c=8$ (for $\chi_\mathrm{top}$) and from $N_c=3$ to $N_c=6$ (for $b_2$), respectively. One thus may check which one of the two different expectations for the large-$N_c$ scalings is supported by the lattice results.
For doing that, we average the results collected in Refs.~\cite{vicari,bonati} (converting $b_2$ to $c_4$) and the more recent results reported in Refs.~\cite{Ce:2015qha,Ce:2016awn}.\footnote{Ref.~\cite{vicari} collects several lattice results from different collaborations. For $N_c > 3$ in the case of $\chi_\mathrm{top}$ and for all $N_c$ in the case of $c_4$, the lattice results are taken from \cite{Lucini:2001ej,DelDebbio:2002xa,Cundy:2002hv,Lucini:2004yh,DElia:2003zne,Giusti:2007tu}; for $N_c = 3$ in the case of $\chi_\mathrm{top}$, \cite{vicari} comprises results from 29 different reports; see table 1 in \cite{vicari} for a complete list of the reports under consideration.} Then we try to fit with functions according to the assumptions in Refs.~\cite{vicari,bonati} (Fit 1: $\chi_\mathrm{top}\sim \mathrm{const.}+\mathcal{O}(N_c^{-2})$ and $c_4\sim \mathcal{O}(N_c^{-2})$) and according to the expectations we found above (Fit 2: $\chi_\mathrm{top}\sim \mathrm{const.}+\mathcal{O}(N_c^{-1})$ and $c_4\sim \mathcal{O}(N_c^{-3})$). The results for the topological susceptibility and the fourth cumulant are given in figs.~\ref{fig:chilattice} and \ref{fig:c4lattice}, respectively, where $\sigma$ is the string tension making $\chi_\mathrm{top}/\sigma^2$ and $c_4/\sigma^2$ dimensionless quantities.

First of all, the lattice calculations are in accordance with the general predictions that the topological susceptibility tends to a finite value in the large-$N_c$ limit, and that $c_4$, though non-zero, is strongly suppressed as $N_c$ increases. However, the lattice results seem to slightly favor the assumptions of Refs.~\cite{vicari,bonati} (Fit~1), though both fits (Fits~1 and 2) for $\chi_\mathrm{top}$ lay well within the error bars, except for $N_c=6$.\footnote{That this data point deviates in comparison to the other data points stems from the fact that not all groups provide results for any SU($N_c$) case and that the results from Refs.~\cite{bonati,DelDebbio:2002xa,Lucini:2004yh} are in general slightly larger in comparision to the results reported in Refs.~\cite{Ce:2016awn,Lucini:2001ej,Cundy:2002hv}: while all Refs. provide data for $N_c=4$, only \cite{Ce:2016awn,Lucini:2001ej,Cundy:2002hv} have results for $N_c=5$, and \cite{bonati,Ce:2016awn,DelDebbio:2002xa,Lucini:2004yh} for $N_c=6$ (thus the slightly shifted data point for the SU(6) case). The latter Refs. are also those that come with the smallest errors. The sole result for SU(8) stems from \cite{Lucini:2004yh}.} For $c_4$ the deviation of Fit~2 is apparent at $N_c=6$. Note, however, that in the case of $c_4$ the lattice results are available for only three different values of $N_c=3,4,6$. Moreover, the respective evaluated values for both $\chi_\mathrm{top}$ and $b_2$ in the respective references fluctuate quite markedly, and the corresponding errors are in some cases relatively large. In addition, the lattice calculations have not yet been performed for truly \emph{large} numbers of colors, while in the previous section the scaling behaviors have been  worked out for $N_c\to\infty$. In the region where $N_c\not\gg 1$, it is clear that the asymptotic expressions in eqs.~\eqref{eq:chitopnloresult} and \eqref{eq:c4nloresult} (or eqs.~\eqref{eq:chitopdeltaresult} and \eqref{eq:c4delta} in the $\delta$-expansion) are not applicable, since the various terms in the full expressions (before sending $N_c$ to infinity) certainly forbid a marked manifestation of the large-$N_c$ scaling for $N_c$ between 3 and 8. Consequently, a naive fit with $\chi_\mathrm{top}=\mathrm{const.}+a/N_c^{(2)}$ and $c_4\sim N_c^{-2(3)}$ is obviously not a suitable choice.

Because of that, we have performed a third fit for both quantities that reflect the structures of the expressions for $\chi_\mathrm{top}$ and $c_4$ given in eqs.~\eqref{eq:chitopdeltaresult} and \eqref{eq:c4delta}, and the results are shown as Fit~3 in figs.\,\ref{fig:chilattice} and \ref{fig:c4lattice}. Now, the fits for the region $N_c\not\gg 1$ is observably better for both quantities, while the large-$N_c$ behavior described by the fit functions is fully in line with the predictions worked out here, i.\,e. $\mathcal{O}(1)+\mathcal{O}(N_c^{-1})$ for $\chi_\mathrm{top}$ and $\mathcal{O}(N_c^{-3})$ for $c_4$.

We also note that several authors agree that for $\chi_\mathrm{top}$ a scaling $\mathcal{O}(1)+\mathcal{O}(N_c^{-1})$ must be correct (see e.\,g. \cite{kaiserleutwyler,shore2007}), in accordance with the results from CHPT obtained here. Note that the constant terms in the lattice fits in fig.~\ref{fig:chilattice} correspond to $\tau$ in eq.~\eqref{eq:chitopnloresult} via
$\tau=\mathrm{const.}\times\sigma^2$ (Fit~1 and 2) and via $\tau=a_3\times\sigma^2$ (Fit~3). The typical value for the string tension $\sqrt{\sigma}$ ranges from 420~MeV to 440~MeV; we take here $\sqrt{\sigma}=440\,\mathrm{MeV}$ as in Ref.~\cite{vicari}. Then these fits lead to 
\begin{equation*}
\tau = \begin{cases}
\left(171.50\pm 1.02 \,\mathrm{MeV}\right)^4\ \mathrm{(Fit~1)}\\
\left(167.70\pm 2.26 \,\mathrm{MeV}\right)^4\ \mathrm{(Fit~2)}\\
\left(172.34\pm 1.81 \,\mathrm{MeV}\right)^4\ \mathrm{(Fit~3)}\ 
\end{cases}
\end{equation*}
which may be compared with the prediction $\tau\approx\left(179\,\mathrm{MeV}\right)^4$ obtained from the Witten--Veneziano formula~\eqref{eq:wittenvenezianofine} derived using leading order CHPT (meson masses adopted from Ref.~\cite{Tanabashi:2018oca}).
The value from Fit~3 should be regarded as our final result.

\section{Physics at \texorpdfstring{{$\theta \sim \pi$}}{Theta sim Pi}}\label{ch:thetasimpi}

\subsection{Previous results for two and three mass-degenerate flavors}

As noted already briefly in the Introduction, $CP$-symmetry, which is theoretically broken in QCD due to the $\theta$-term, would be restored not only if $\theta=0$, but also if $\theta=\pi$. However, Di Vecchia and Veneziano~\cite{vecchiaveneziano1980}, and Witten~\cite{witten1980} showed already that at $\theta=\pi$ it is possible that more than one vacuum may appear, from which none may turn out to be $CP$-noninvariant meaning that $CP$ would be spontaneously broken. 
For the case of two or three mass-degenerate flavors, this spontaneous $CP$-symmetry breaking known as Dashen's phenomenon/mechanism~\cite{dashen} indeed occurs as has been shown already by the above mentioned authors.\footnote{We note that the appearance of spontaneous CP symmetry breaking and two degenerate vacua at $\theta=\pi$ may be explained as a consequence of a mixed 't Hooft anomaly at $\theta=\pi$ between the global flavor symmetry and the CP symmetry; cf. Refs. \cite{Gaiotto:2017yup,Gaiotto:2017tne}.}

The situation at $\theta\sim\pi$ has been studied in more detail by Smilga~\cite{smilga} considering the cases of two and three mass-degenerate flavors in the framework of CHPT. He demonstrated that in both cases the vacuum at $\theta=\pi$ is indeed degenerate and there exists a region around $\theta=\pi$, which may be denoted as $[\pi-\epsilon, \pi+\epsilon]$, in which two local minima coexist, one being the stable, true vacuum, the other being a metastable, false vacuum. For $N_f=3$ flavors these phenomena are readily worked out by minimizing the leading order effective potential. This is, however, not applicable for $N_f=2$ flavors, since the leading order potential at $\theta=\pi$ does not depend on the choice of the matrix $U$ in this case. So in order to recover the expected behavior also for $N_f=2$ flavors, one has to consider the next-to-leading order Lagrangian (this insight goes back to the findings of Creutz in Ref.~\cite{creutz}).

There are, however, large quantitative differences between the two cases: while the width of the region for two local minima in the case of three flavors is given by $\epsilon=\pi/2$, the width of the region of two coexisting local minima in the $N_f=2$ case is much smaller:
\begin{equation}\label{eq:regionofcoexistingminimanfgleich2}
\epsilon=\frac{8l_7\Sigma m}{F_\pi^4}\approx\frac{2}{3}\frac{M_\pi^2}{M_\eta^2}\approx 0.04\, ,
\end{equation}
where we have inserted the Gell-Mann--Oakes--Renner (GMOR) relation and use $l_7=F_\pi^2/(6M_\eta^2)$ for the low-energy constant  $l_7$~\cite{gasserleutwyler1,smilga}. The same is true for the domain wall tension:
\begin{align}
T_W^{N_f=2} & = M_\pi^2F_\pi\sqrt{8l_7}\approx 0.2\times M_\pi^2F_\pi\,, \nonumber \\
T_W^{N_f=3} & = M_\pi^2F_\pi 3\sqrt{2}\left(1-\frac{\pi}{3\sqrt{3}} \right)\approx 1.7\times M_\pi^2F_\pi\, . \label{eq:smilgaswalltension}
\end{align}
In the following, we  will derive the corresponding properties for the $2+1$ flavor case. It will be demonstrated that the results for this particular case are qualitatively compatible with the previous findings of Smilga.

\subsection{Physics at \texorpdfstring{$\theta \sim \pi$}{Theta sim Pi} for \texorpdfstring{$2+1$}{2+1} light flavors}

\subsubsection{Stationary points and the region of two coexisting local minima}

\begin{figure}[tb]
\centering
\includegraphics[width=0.7\textwidth]{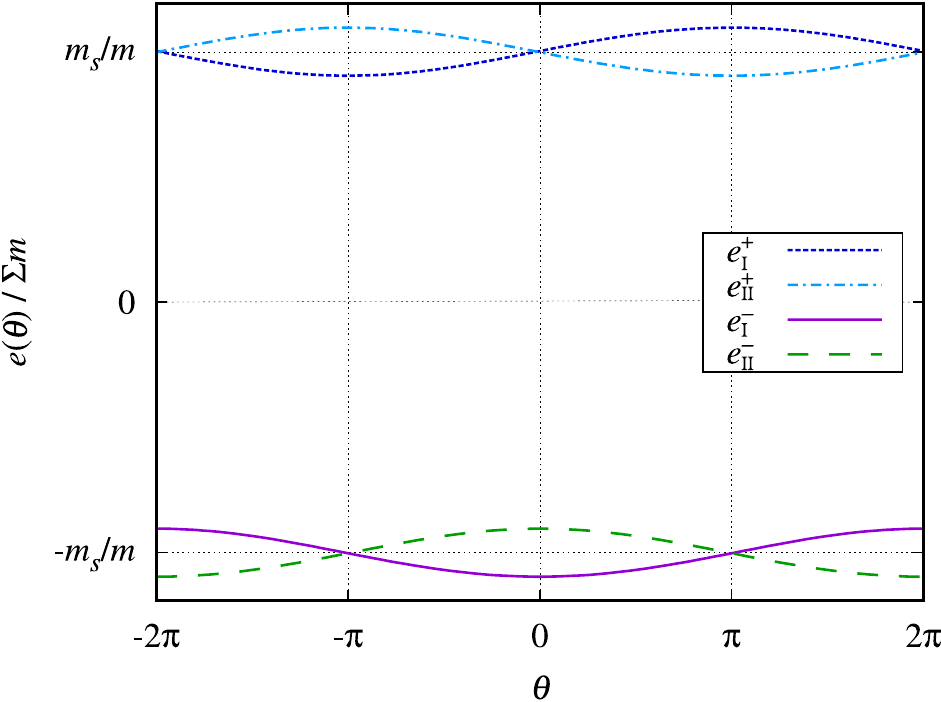}
\caption{Basic stationary points of $V_\pm(\theta,\alpha)$. For details, see the text.}
\label{fig:stationarypointsLO}
\end{figure}

Before writing down the effective chiral Lagrangian, let us recall that with a suitable choice of a global U(1)$_A$ transformation one may remove the $\theta$-term from the Lagrangian, and the $\theta$-vacuum angle will be shifted to the phase of the quark mass matrix $\mathcal{M} \to  \mathcal{M}_\theta=e^{i{\theta}/{N_f}}\mathcal{M}$, i.\,e., for the classical action the equation
\[
S\{\theta, \mathcal{M}\} = S\{0, \mathcal{M}_\theta\}
\]
is valid. 
This  results in the leading order Lagrangian
\begin{equation}\label{eq:lagranthetasimpi}
\mathcal{L} = \frac{F^2}{4}\operatorname{Tr}\left[\partial_\mu U^\dagger \partial^\mu U\right]+\Sigma\operatorname{Re}\operatorname{Tr}\left[e^{i\frac{\theta}{3}}\mathcal{M} U^\dagger\right] .
\end{equation}
The mass matrix for $m_u = m_d \equiv m$ and $0<m\ll m_s$ will be considered as being real and diagonal, i.\,e. $\mathcal{M}=\operatorname{diag}\left(m,m,m_s\right)$. Moreover, any unitary matrix can be transformed into the diagonal form $U=\operatorname{diag}\left(e^{i\alpha},e^{i\beta},e^{-i(\alpha+\beta)}\right)$, so the potential of the Lagrangian \eqref{eq:lagranthetasimpi} may be written as
\begin{equation}\label{eq:potthetasimpi}
V(\alpha,\beta,\theta)=-\Sigma m\left\{\cos\left(\frac{\theta}{3}-\alpha\right)+\cos\left(\frac{\theta}{3}-\beta\right)+\frac{m_s}{m}\cos\left(\frac{\theta}{3}+\alpha+\beta\right)\right\} .
\end{equation}
Stationary points of this potential are found by the condition $\partial V/\partial\alpha=\partial V/\partial\beta=0$. Obviously, the solutions for $\alpha, \beta$ are symmetric under the exchange $\alpha \leftrightarrow \beta$. So one may, for instance, determine $\beta$ in terms of $\alpha$ first, and find the corresponding solutions for $\alpha$ afterwards. Consider thus first $\partial V/\partial\alpha=0$ resulting in
\[
\sin\left(\frac{\theta}{3}+\alpha+\beta\right) = \frac{m}{m_s}\sin\left(\frac{\theta}{3}-\alpha\right),
\]
which gives two solutions for $\beta$:\footnote{All solutions for $\alpha$ and $\beta$ that will be given in the following  are obviously periodic with period $2\pi$. We will, however, refrain from writing explicitly the additional term $+2\pi n$, $n\in \mathbb{Z}$, that would take this into account.}
\begin{align}
\beta_{-}&=\arcsin\left[\frac{m}{m_s}\sin\left(\frac{\theta}{3}-\alpha\right)\right]-\frac{\theta}{3}-\alpha\,, \label{eq:betaminus}\\
\beta_{+} & = \beta_{-}+\pi\, . \nonumber
\end{align}
The choice of the subscript $\pm$ will become clear when inserting these solutions into the potential \eqref{eq:potthetasimpi},
\begin{align*}
V_\pm (\alpha,\theta)= &\Sigma m\Biggl\{\biggr.-\cos\left(\frac{\theta}{3}-\alpha\right)\pm\cos\left(\frac{2\theta}{3}+\alpha-\arcsin\left[\frac{m}{m_s}\sin\left(\frac{\theta}{3}-\alpha\right)\right]\right)\\ &\pm\frac{m_s}{m}\sqrt{1-\left(\frac{m}{m_s}\right)^2\sin^2\left(\frac{\theta}{3}-\alpha\right)}\biggl.\Biggr\}\, .
\end{align*}
Observe that the last term is approximately constant, since $\left({m}/{m_s}\right)^2$ may be approximated by zero. This almost constant term $\sim \pm{m_s}/{m}\gg 1$ implies hence that solutions for $\beta_{+}$ will always have positive energy densities, while solutions for $\beta_{-}$ will always have negative energy densities. Vacuum solutions are thus to be expected for $V_{-}$. With the condition $\partial V_{-}/\partial \alpha=0$ and the approximation $1+\frac{m}{m_s}\cos\left(\frac{\theta}{3}-\alpha\right)\approx 1$ one finds
\begin{equation}\label{eq:vminusstationarypointsequation}
\sin\left(\frac{\theta}{3}-\alpha\right)=\sin\left(\frac{2\theta}{3}+\alpha-\arcsin\left[\frac{m}{m_s}\sin\left(\frac{\theta}{3}-\alpha\right)\right]\right) .
\end{equation}
Approximating ${m}/{m_s}\approx 0$, this gives
\begin{equation}\label{eq:vminusstationarypointsequation2}
\sin\left(\frac{\theta}{3}-\alpha\right)=\sin\left(\frac{2\theta}{3}+\alpha\right) ,
\end{equation}
which is solved by
\begin{align*}
\alpha_\mathrm{I}^{-} & = -\frac{\theta}{6}\,,\\
\alpha_\mathrm{II}^{-} & = \pi-\frac{\theta}{6}
\end{align*}
with the corresponding energy densities
\begin{align*}
e_\mathrm{I}^{-}(\theta) & = \Sigma m \left\{-\cos\frac{\theta}{2} - \cos\left(\frac{\theta}{2}-\arcsin\left[\frac{m}{m_s}\sin\frac{\theta}{2}\right]\right)-\frac{m_s}{m}\right\},\\
e_\mathrm{II}^{-}(\theta) & =\Sigma m \left\{\cos\frac{\theta}{2} + \cos\left(\frac{\theta}{2}+\arcsin\left[\frac{m}{m_s}\sin\frac{\theta}{2}\right]\right)-\frac{m_s}{m}\right\}  .
\end{align*}
Following the same strategy, one can determine also the pertinent positive solutions
\begin{align*}
\alpha_\mathrm{I}^{+} & = \frac{3\pi}{2}-\frac{\theta}{6}\,,\\
\alpha_\mathrm{II}^{+} & = \frac{\pi}{2}-\frac{\theta}{6}
\end{align*}
with energy densities
\begin{align*}
e_\mathrm{I}^{+}(\theta) & = \Sigma m \left\{\sin\frac{\theta}{2} + \sin\left(\frac{\theta}{2}-\arcsin\left[\frac{m}{m_s}\cos\frac{\theta}{2}\right]\right)+\frac{m_s}{m}\right\},\\
e_\mathrm{II}^{+}(\theta) & =\Sigma m \left\{-\sin\frac{\theta}{2} - \sin\left(\frac{\theta}{2}+\arcsin\left[\frac{m}{m_s}\cos\frac{\theta}{2}\right]\right)+\frac{m_s}{m}\right\}  .
\end{align*}
These basic solutions are shown in fig.~\ref{fig:stationarypointsLO} which may be compared with the corresponding fig.\,1 in \cite{smilga} for $N_f=3$, $m_u=m_d=m_s=m$. Obviously the negative and positive solutions are separated by the shift $\pm \Sigma m_s$ (corresponding to a shift $\pm m_s/m$ in fig.~\ref{fig:stationarypointsLO}). If $m_s\to m$, the curves would move closer together. Observe furthermore that if $m_s\to\infty$ the solutions $e_\mathrm{I/II}^\pm$ become independent of $\theta$ and $m$. In fact, the potential $V_\pm$ becomes constant as well, $V_\pm=\pm \Sigma m_s$, so it is independent of the choice of $\alpha$, which is in accordance with the observation that for the case $N_f=2$ the potential is independent of the choice of $U$ at leading order.

\begin{figure}[tb]
\centering
\includegraphics[width=0.7\textwidth]{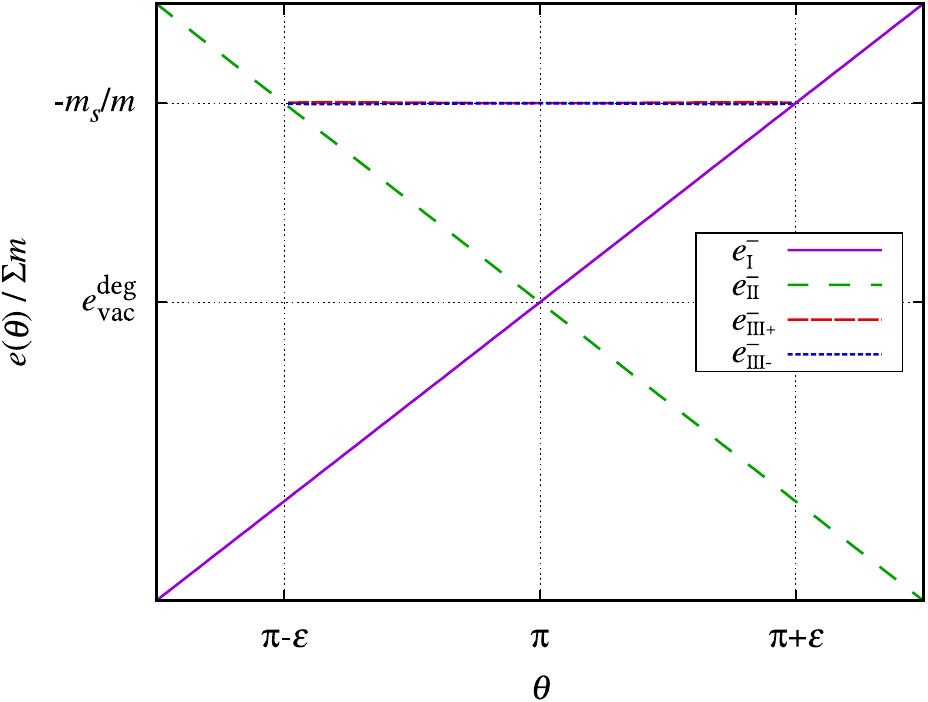}
\caption{Stationary points of $V_{-}(\theta,\alpha)$ at $\theta\sim\pi$.}
\label{fig:stationarypointsLO2}
\end{figure}

However, before testing which of these points describe extrema or saddle points, respectively (which can of course already be deduced  from fig.~\ref{fig:stationarypointsLO}---at least roughly), one should note that eq.~\eqref{eq:vminusstationarypointsequation2} at $\theta=\pi$ is solved by \emph{any} $\alpha$, so obviously the approximation assumed above is not applicable for this case. To see what really is going on in the region $\theta\sim\pi$, which is actually the case we are interested in, we have to seek for another solution valid in this region. In order to achieve that, one may introduce $\theta^\prime=\theta-\pi$ for which we will assume $0\leqslant |\theta^\prime|\ll 1$. Inserting this into eq.~\eqref{eq:vminusstationarypointsequation}, using $\cos{x}\approx 1$ and $\sin{x}\approx x$ for $0\leqslant |x|\ll 1$, and neglecting terms $\sim \theta^\prime{m}/{m_s}$, $\sim {\theta^\prime}^2$, and $\sim \left({m}/{m_s}\right)^2$, one gets
\[
0=\theta^\prime\left[\cos\alpha+\sqrt{3}\sin\alpha\right]-\frac{m}{2m_s}\left[\sin{2\alpha}+\sqrt{3}\cos{2\alpha}\right] .
\]
This equation gives two additional solutions
\[
\alpha_{\mathrm{III}\pm}^{-}=-2\arctan\left[\frac{\frac{m}{m_s}\pm 2\sqrt{\left(\frac{m}{m_s}\right)^2-{\theta^\prime}^2}}{\sqrt{3}\frac{m}{m_s}+2\theta^\prime}\right] .
\]
Because of the expression in the square root, these solutions are real only for $|\theta^\prime|\leqslant\frac{m}{m_s}$.\footnote{Similar solutions can be found also for $V_{+}$, for which the corresponding condition, this time at $\theta=0$, in the approximation ${m}/{m_s}\approx 0$ is solved by any $\alpha$, but we will not give it here explicitly as we are interested in particular in the vacuum states. These additional solutions appear in the region at $\theta\sim 0$ where the solutions $\alpha_\mathrm{I/II}^{+}$ describe two coexisting saddle points.} Consequently, this suggests that it is within this particular region, i.\,e. $[\pi-{m}/{m_s},\pi+{m}/{m_s}]$, that two local minima coexist.
In order to establish this, one has to calculate $\partial^2 V/\partial^2\alpha$, $\partial^2 V/\partial^2\beta$, and $\partial^2 V/\partial\alpha\partial\beta$ and check their behavior for the solutions just found:
\begin{align*}
\frac{\partial^2 V}{\partial^2\alpha} &= \Sigma m\left\{\cos\left(\frac{\theta}{3}-\alpha\right)+\frac{m_s}{m}\cos\left(\frac{\theta}{3}+\alpha+\beta\right)\right\},\\
\frac{\partial^2 V}{\partial^2\beta} &= \Sigma m\left\{\cos\left(\frac{\theta}{3}-\beta\right)+\frac{m_s}{m}\cos\left(\frac{\theta}{3}+\alpha+\beta\right)\right\},\\
\frac{\partial^2 V}{\partial\alpha\partial\beta} &= \Sigma m \frac{m_s}{m}\cos\left(\frac{\theta}{3}+\alpha+\beta\right).
\end{align*}
The result is shown in table \ref{tab:stationary}, where $\Delta^2$ has been defined as
\[
\Delta^2 :=\frac{\partial^2 V}{\partial^2\alpha}\frac{\partial^2 V}{\partial^2\beta}-\left(\frac{\partial^2 V}{\partial\alpha\partial\beta}\right)^2\ .
\]
\begin{table}\centering
\begin{tabular}{lll}
\hline
\\[-1em]
$\alpha_\mathrm{I}^{+}$: & $\frac{\partial^2 V}{\partial\alpha^2},\frac{\partial^2 V}{\partial\beta^2}<0\ \forall\ \theta$ & \\[2mm]
& (a) $\Delta^2<0$ for $\theta\in [0,\frac{m}{m_s})$ and $\theta\in (2\pi-\frac{m}{m_s},2\pi]$ & $\Rightarrow$ saddle point\\[2mm]
& (b) $\Delta^2>0$ for $\theta\in (\frac{m}{m_s},2\pi-\frac{m}{m_s})$ & $\Rightarrow$ absolute maximum\\[2mm] \hline 
\\[-1em]
$\alpha_\mathrm{II}^{+}$: & $\frac{\partial^2 V}{\partial\alpha^2},\frac{\partial^2 V}{\partial\beta^2}<0\ \forall\ \theta$ & \\[2mm]
& $\Delta^2<0\ \forall\ \theta$ & $\Rightarrow$ saddle point\\[2mm] \hline 
\\[-1em]
$\alpha_\mathrm{I}^{-}$: & $\frac{\partial^2 V}{\partial\alpha^2},\frac{\partial^2 V}{\partial\beta^2}>0\ \forall\ \theta$ & \\[2mm]
& (a) $\Delta^2>0$ for $\theta\in [0,\pi+\frac{m}{m_s})$ & $\Rightarrow$ local minimum\\[2mm]
& (b) $\Delta^2<0$ for $\theta\in (\pi+\frac{m}{m_s},2\pi]$ & $\Rightarrow$ saddle point\\[2mm] \hline 
\\[-1em]
$\alpha_\mathrm{II}^{-}$ & $\frac{\partial^2 V}{\partial\alpha^2},\frac{\partial^2 V}{\partial\beta^2}>0\ \forall\ \theta$ & \\[2mm]
& (a) $\Delta^2<0$ for $\theta\in [0,\pi-\frac{m}{m_s})$ & $\Rightarrow$ saddle point\\[2mm]
& (b) $\Delta^2>0$ for $\theta\in (\pi-\frac{m}{m_s},2\pi]$ & $\Rightarrow$ local minimum\\[2mm] \hline 
\\[-1em]
$\alpha_\mathrm{III\pm}^{-}$: & $\frac{\partial^2 V}{\partial\alpha^2},\frac{\partial^2 V}{\partial\beta^2}>0$ for $\theta\in [\pi-\frac{m}{m_s},\pi+\frac{m}{m_s}]$ & \\[2mm]
& $\Delta^2<0$ for $\theta\in [\pi-\frac{m}{m_s},\pi+\frac{m}{m_s}]$ & $\Rightarrow$ saddle point\\[2mm] \hline
\end{tabular}
\caption{Classification of the stationary points of $V(\alpha,\beta,\theta)$.}
\label{tab:stationary}
\end{table}
This means for the vacuum states, that
\begin{enumerate}
\item in the region $[0,\pi-\frac{m}{m_s})$, $\alpha_\mathrm{I}^{-}$ is the absolute minimum describing the vacuum state with an energy density $e_\mathrm{vac}(\theta)=e_\mathrm{I}^{-}(\theta)$;
\item in the region $(\pi-\frac{m}{m_s},\pi)$, two local minima coexist with $\alpha_\mathrm{I}^{-}$ being the stable vacuum, while $\alpha_\mathrm{II}^{-}$ describes a metastable vacuum, while the solutions $\alpha_\mathrm{III\pm}^{-}$ describe two saddle points appearing between these minima;
\item at $\theta=\pi$, the two local minima describe a degenerate vacuum with an energy density $e_\mathrm{vac}^\mathrm{deg}=e_\mathrm{I}^{-}(\pi)=e_\mathrm{II}^{-}(\pi)=-\Sigma m\left(\frac{m}{m_s}+\frac{m_s}{m}\right)$;
\item in the region $(\pi,\pi+\frac{m}{m_s})$, two local minima coexist with $\alpha_\mathrm{II}^{-}$ being the stable vacuum, while $\alpha_\mathrm{I}^{-}$ now describes the metastable vacuum;
\item in the region $(\pi+\frac{m}{m_s},2\pi]$, the saddle points $\alpha_\mathrm{III\pm}^{-}$ disappear and  $\alpha_\mathrm{II}^{-}$ is the absolute minimum describing the vacuum state, while $\alpha_\mathrm{I}^{-}$ is a saddle point.
\end{enumerate} 
This behavior is depicted in fig.~\ref{fig:stationarypointsLO2} for the crucial region around $\theta=\pi$.

Actually, the region of two local minima is not exactly given by $[\pi-\frac{m}{m_s},\pi+\frac{m}{m_s}]$. Let this region be $[\pi-\epsilon,\pi+\epsilon]$ (fig.\,\ref{fig:stationarypointsLO2}). 
The quantity $\epsilon$ may be determined by calculating the position where $e_\mathrm{I/II}^{-}/(\Sigma m)$ passes $-{m_s}/{m}$ (it is exactly below this threshold where two local minima coexist, see fig.\,\ref{fig:stationarypointsLO2}). By setting $e_\mathrm{I/II}^{-}/(\Sigma m)=-{m_s}/{m}$, one gets
\[
\sin\frac{\epsilon}{2}+\sin\left(\frac{\epsilon}{2}-\arcsin\left[\frac{m}{m_s}\cos\frac{\epsilon}{2}\right]\right)=0\, .
\]
For $\cos(\epsilon/2)\approx 1$, this is solved by
\begin{equation}\label{eq:epsilonregion2}
\epsilon = \arcsin\left(\frac{m}{m_s}\right) ,
\end{equation}
which gives for small $m/m_s$ the result $\epsilon\approx m/m_s$. The width of the region of two coexisting minima is hence comparable to the width of the region obtained for $N_f=2$ flavors (see eq.\,\eqref{eq:regionofcoexistingminimanfgleich2}). Although their origin seems different, they can be matched by observing eq.~\eqref{eq:regionofcoexistingminimanfgleich2} gives $2M_\pi^2/(3M_\eta^2)\approx m/m_s$ using the GMOR relation. The reason is that the $l_7$ term in the SU(2) case is saturated by the $\eta$ meson which is included explicitly as a pseudo-Goldstone boson in the $2+1$ flavor case.

In summary, the region of two coexisting local minima increases as $m_s/m$ decreases, while at the same time the positive and negative energy solutions move closer together. 
The solutions presented here are of course only valid as long as $m/m_s$ can be considered as being ``small''. For $m/m_s \not\ll 1$ the solutions will differ and change into the solutions found by Smilga for $m=m_s$, where $\epsilon={\pi}/{2}$ (this can also be found from eq.\,\eqref{eq:epsilonregion2} setting $m=m_s$).

\subsubsection{The domain wall at \texorpdfstring{$\theta=\pi$}{Theta Pi} and the wall tension}

Between the two degenerate vacua at $\theta=\pi$, i.\,e. the minima at $\alpha_\mathrm{I}^{-}$ and $\alpha_\mathrm{II}^{-}$, appears a domain wall, whose total energy is infinite. In this section, the profile of this wall in one space direction will be calculated as well as its surface tension $T_W$. Having calculated the wall tension, one can estimate the decay rate $\Gamma$ of the false vacuum in the vicinity of $\theta=\pi$.

\begin{figure}
\centering
\includegraphics[width=0.7\textwidth]{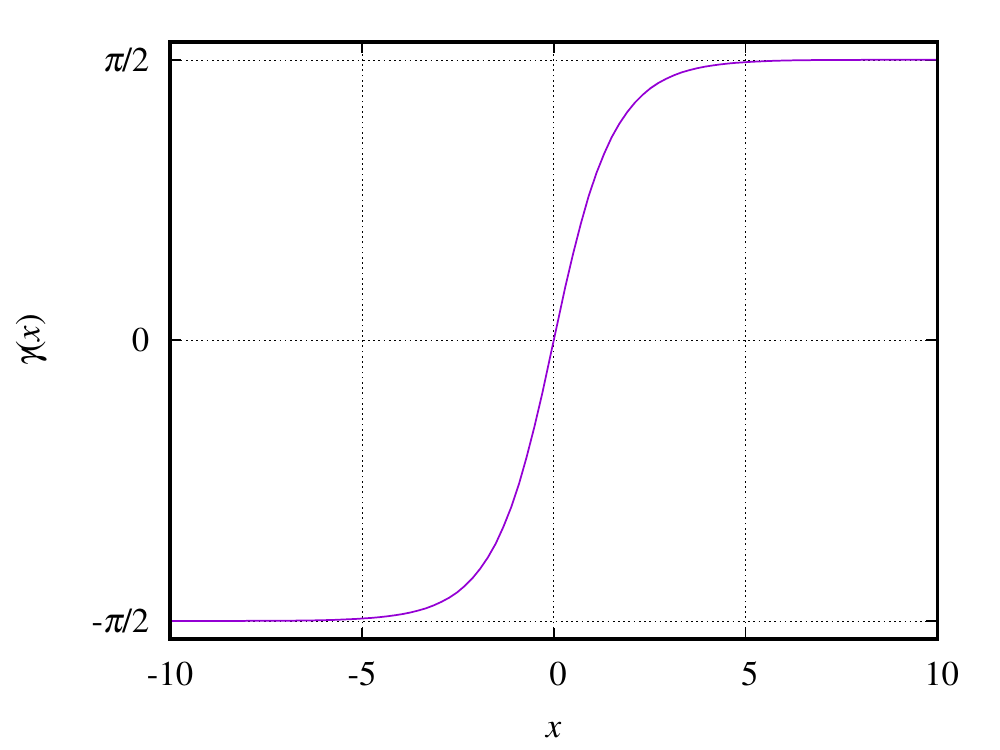}
\caption{Transition between vacuum~II with $\gamma(z)=-{\pi}/{2}$ and vacuum~I with $\gamma(z)={\pi}/{2}$ using the substitution $x=M_\pi\sqrt{\frac{m}{m_s}}z$.}
\label{fig:domainwall}
\end{figure}

Let $\mathcal{A}=\int \mathrm{d}x\,\mathrm{d}y$ be the infinite wall area extended in the $xy$-plane. The field configuration of the domain wall then only depends on the $z$-direction (the wall is static). Inserting the expression for $\beta_{-}$ from eq.\,\eqref{eq:betaminus} and $\theta=\pi$ gives
\begin{equation*}
U(z)=\operatorname{diag}\left(e^{i\alpha(z)}, e^{i\left(\arcsin\left[\frac{m}{m_s}\sin\left(\frac{\pi}{3}-\alpha(z)\right) \right] -\frac{\pi}{3} -\alpha(z) \right)}, e^{-i\left(\arcsin\left[\frac{m}{m_s}\sin\left(\frac{\pi}{3}-\alpha(z)\right) \right] -\frac{\pi}{3}\right)} \right) .
\end{equation*}
Assume that at one end of the universe, at $z=-\infty$, the system is in one of the two degenerate vacuum states, say at $\alpha_\mathrm{II}^{-}$, and at the other end of the universe, the system is in the other vacuum state $\alpha_\mathrm{I}^{-}$. Then the boundary conditions are
\begin{align*}
\alpha(z\to -\infty) & =\alpha_\mathrm{II}^{-} (\theta=\pi)=\frac{5\pi}{6}\,,\\
\alpha(z\to +\infty) & =\alpha_\mathrm{I}^{-}(\theta=\pi)=-\frac{\pi}{6} \, .
\end{align*}
The wall tension and the field configuration describing the profile of the wall are found by considering and minimizing
\[
E_W - E_\mathrm{vac} = \mathcal{A}\,T_W\, ,
\]
where $E_W$ is the total energy of the wall and 
\[
E_\mathrm{vac}=\mathcal{A}\int\mathrm{d}z\, e_\mathrm{vac}\,.
\]
Note that all integrals in $z$-direction that are given in this section have of course to be integrated from $-\infty$ to $+\infty$, which we will not write explicitly. The vacuum energy density is
\begin{equation*}
e_\mathrm{vac}=e_\mathrm{I/II}^{-}(\theta=\pi)=-\Sigma m \left(\frac{m}{m_s}+\frac{m_s}{m}\right)  .
\end{equation*}
With the Lagrangian \eqref{eq:lagranthetasimpi} the wall tension is hence given by (using $\alpha^\prime(z)={\partial \alpha(z)}/{\partial z}$)
\begin{align*}
T_W & = \frac{E_\mathrm{W}}{\mathcal{A}}-\int\mathrm{d}z\, e_\mathrm{vac}\\
& = \int\mathrm{d}z\, \left\{\frac{F^2}{4}\operatorname{Tr}\left[\partial_z U^\dagger \partial_z U\right]-\Sigma\operatorname{Re}\operatorname{Tr}\left[e^{i\frac{\theta}{3}}\mathcal{M} U^\dagger\right]-e_\mathrm{vac}  \right\}\\
& = \int\mathrm{d}z\, \Biggl\{\Biggr.\frac{F^2}{2} {\alpha^\prime}^2(z) \left(1+\frac{m}{m_s}\cos\left(\frac{\pi}{3}-\alpha(z)\right)\right)-\Sigma m\biggl(\biggr. \cos\left(\frac{\pi}{3}-\alpha(z)\right)\\
& \qquad\quad  + \cos\left(\frac{2\pi}{3}-\arcsin\left[\frac{m}{m_s}\sin\left(\frac{\pi}{3}-\alpha (z)\right)\right] + \alpha(z)\right) -\frac{m}{m_s} \biggl.\biggr) \Biggl.\Biggr\}\, ,
\end{align*}
where terms $\sim\left({m}/{m_s} \right)^2$ have been omitted. Let
\[
\gamma(z)=\frac{\pi}{3}-\alpha(z)\,,
\]
so that the boundary conditions are
\begin{equation}\label{eq:boundarycondgamma}
\gamma(z\to \pm\infty) = \pm \frac{\pi}{2}\, .
\end{equation}
Then
\begin{equation}\label{eq:walltensionwithgamma}
T_W = F^2 \int \mathrm{d}z\, \left\{\frac{1}{2}{\gamma^\prime}^2(z)\left(1+\frac{m}{m_s}\cos\gamma(z) \right) + \frac{M_\pi^2}{2}\frac{m}{m_s}\cos^2\gamma(z) \right\} ,
\end{equation}
where the GMOR relation has been used.
Minimizing this expression yields the second-order nonlinear differential equation
\begin{equation}\label{eq:secondorderdiffeq}
\gamma^{\prime\prime}(z)=-\frac{M_\pi^2 m}{m_s}\frac{\sin\gamma(z) \cos\gamma(z)}{1+\frac{m}{m_s}\cos\gamma(z)}\ ,
\end{equation}
which can be solved by approximating $1+\frac{m}{m_s}\cos\gamma(z)\approx 1$.\footnote{As it turns out, $\cos\gamma(z)\sim {m}/{m_s}$, see eq.\,\eqref{eq:diffeqsolutiongamma}, so this is indeed an appropriate approximation.} Multiplying both sides with $2\gamma^\prime(z)$, eq.~\eqref{eq:secondorderdiffeq} can be transformed into a first-order differential equation
\begin{equation}\label{eq:firstorderdiffgamma}
\gamma^\prime(z)=M_\pi\sqrt{\frac{m}{m_s}}\cos\gamma(z)\ ,
\end{equation}
which is solved by
\begin{equation}\label{eq:diffeqsolutiongamma}
\gamma(z)=2\arctan\left[\tanh\left(\frac{M_\pi}{2}\sqrt{\frac{m}{m_s}}z-z_0\right)\right] .
\end{equation}
This equation indeed fulfills the boundary conditions in eq.~\eqref{eq:boundarycondgamma}:
\[
\gamma(z\to\pm\infty) = 2\arctan(\pm 1) = \pm\frac{\pi}{2}\, .
\]
The constant $z_0$ denotes the point where $\gamma(z)$ crosses the $z$-axis and may hence be set to zero, $z_0=0$. The profile of this field configuration is shown in fig.~\ref{fig:domainwall}, where the dimensionless parameter $x=M_\pi\sqrt{\frac{m}{m_s}}z$ has been substituted. The transition from vacuum~II to vacuum~I appears roughly between $x=-5$ and $x=5$ corresponding to a width $\Delta z =\frac{10}{M_\pi}\sqrt{\frac{m_s}{m}}\approx 75\,\mathrm{fm}$. Note that at $\gamma(z=0)=0$, corresponding to $\alpha(z=0)=\frac{\pi}{3}=\alpha_\mathrm{III-}^{-}(\theta=\pi)$, meaning that the solution presented here describes a transition in which $\alpha(z)$ passes the saddle point III$-$.

Inserting $\gamma^\prime$ from equation \eqref{eq:firstorderdiffgamma} back into the expression \eqref{eq:walltensionwithgamma} for the wall tension and using the solution just found yields
\begin{align}
T_W & = F^2M_\pi^2\frac{m}{m_s} \int\limits_{-\infty}^{\infty}\mathrm{d}z\, \left(\frac{1-\tanh^2\left(\frac{M_\pi}{2}\sqrt{\frac{m}{m_s}}z\right)}{1+\tanh^2\left(\frac{M_\pi}{2}\sqrt{\frac{m}{m_s}}z\right)} \right)^2 \nonumber\\
& = 2F^2 M_\pi \sqrt{\frac{m}{m_s}}\, , \label{eq:walltensionfinal}
\end{align}
which is $\sqrt{2}$ times  Smilga's result for $N_f=2$ mass-degenerate flavors given above in eq.~\eqref{eq:smilgaswalltension} (using the GMOR relation and the $\eta$ saturation for $l_7$). 
%
For $m=m_s$ the $2+1$ flavor result roughly agrees with Smilga's $N_f=3$ result, the deviation of the numerical value of the prefactor is certainly a consequence of the approximations applied during the derivation of the wall tension given in eq.~\eqref{eq:walltensionfinal}, which are indeed not valid for $m\not\ll m_s$.

The values of the energies of two non-degenerate vacua in the vicinity of $\theta=\pi$ are separated by
\[
\Delta e \approx 2\Sigma m|\theta^\prime|=F^2M_\pi^2|\theta^\prime|\, ,
\]
where $\theta^\prime$ is given by $\theta^\prime=\theta-\pi$, $0<|\theta^\prime|<{m}/{m_s}\ll 1$. Inserting this and the domain wall tension \eqref{eq:walltensionfinal} into eq.~(10) in Ref.~\cite{smilga}, the decay rate of a metastable vacuum can be estimated as 
\[
\Gamma \sim \exp\left(-6^3\pi^2 \frac{F^2}{M_\pi^2}\left(\frac{m}{m_s}\right)^2 \frac{1}{|\theta^\prime|^3} \right) .
\]
Smilga's $N_f=3$ result reads
\[
\Gamma^{N_f=3} \sim \exp\left(-0.8\times 6^3\pi^2 \frac{F^2}{M_\pi^2} \frac{1}{|\theta^\prime|^3} \right),
\]
meaning that the lifetime of a metastable vacuum in the $N_f=3$ mass-degenerate flavor case is much longer than that in the $2+1$ flavor case, where the additional factor $\left({m}/{m_s}\right)^2$ appears.

\subsection{Discussion of the results}

In the previous section, the stationary points and vacuum properties of a theory consisting of $2+1$ light quark flavors have been worked out  to leading order in CHPT considering $m\ll m_s$. As in the cases of two and three mass-degenerate quarks, the vacuum is unique at $\theta\sim 0$ but twofold degenerate at $\theta=\pi$. Around $\theta=\pi$ there is a region of two local minima. In particular, it has been shown that in the $2+1$ flavor case, the maxima and minima are separated by an energy shift $\pm \Sigma m_s$ and that only in the region of two coexisting minima two additional saddle points $\alpha_\mathrm{III\pm}^{-}$ show up.

The main results for the case of $2+1$ flavors are
\begin{itemize}
\item the region of two local minima: $\left[\pi-\arcsin\left(\frac{m}{m_s}\right), \pi+\arcsin\left(\frac{m}{m_s}\right)\right]$,
\item the domain wall tension: $T_W= 2F^2 M_\pi \sqrt{\frac{m}{m_s}} $,
\item the decay rate of metastable vacuum: $\Gamma \sim \exp\left(-6^3\pi^2 \frac{F^2}{M_\pi^2}\left(\frac{m}{m_s}\right)^2 \frac{1}{|\theta^\prime|^3} \right)$.
\end{itemize}
The $N_f=3$ case may be obtained from the $N_f=2+1$ case by taking the limit $m_s\to m$ (qualitatively, since during the derivation contributions $\sim (m/m_s)^2$ were neglected). Indeed, the region of two local minima increases for $m_s\to m$ while the minimum (or minima for $\theta\in \left[\pi-\arcsin\left(\frac{m}{m_s}\right), \pi+\arcsin\left(\frac{m}{m_s}\right)\right]$) and the maximum of the potential move closer together. In the case of $N_f=3$ flavors the energies of the extrema finally meet each other at some particular points, as can be seen in fig.~1 in Smilga's paper. For $m=m_s$, the expressions for the wall tension and the decay rate agree with the expressions for $N_f=3$ mass-degenerate flavors up to a slightly different prefactor---certainly a consequence of the approximations applied.

On the other hand, the $N_f=2$ mass-degenerate flavor result should be obtained by taking the limit $m_s\to \infty$, i.\,e. by decoupling the strange quark from the theory. It is easy to see that in the limit $m_s\to \infty$ the region of two coexisting minima vanishes as well as the wall tension, meaning that one has to consider, just as in the case of $N_f=2$ flavors, the NLO terms $\sim L_3$ and $\sim L_7$ in order to reproduce the $N_f=2$ case. But this turns out to be a rather difficult task in the $2+1$ flavor case. Consider only the $L_7$-term neglecting the term $\sim L_3$ (as has been done by Smilga with respect to the corresponding SU(2) terms $\sim l_3$ and $\sim l_7$):
\begin{equation}\label{eq:l7term}
\mathcal{L}_{L_7} = -4 L_7 \left(\operatorname{Im}\operatorname{Tr}\left[\chi U^\dagger\right]\right)^2\, .
\end{equation}
Inserting $\chi=2({\Sigma}/{F^2})e^{i\theta/3}\mathcal{M}$, with $\mathcal{M}=\operatorname{diag}\left(m,m,m_s\right)$, the GMOR relation $M_\pi^2=2\Sigma m/F^2$ and $L_7\approx -F^2/(24M_{\eta^\prime}^2)$ \cite{gasserleutwyler2}\footnote{The relation is actually given by $L_7 = -\gamma^2 F^2/(48M_{\eta^\prime})$ \cite{gasserleutwyler2}, but since the value of $L_7$ has been determined as being $L_7\approx 0.4\times 10^{-3}$, one can anticipate that roughly $\gamma\approx \sqrt{2}$, which is a fairly good approximation for the subsequent discussion.}, the potential of the $L_7$-term \eqref{eq:l7term} including the leading order contributions may be written as
\begin{align*}
V(\alpha,\beta,\theta) = & - \frac{M_\pi^2 F^2 m_s}{2m} \Biggl\{\Biggr. \frac{m}{m_s}\cos\left(\frac{\theta}{3}-\alpha\right)+\frac{m}{m_s}\cos\left(\frac{\theta}{3}-\beta\right)+\cos\left(\frac{\theta}{3}+\alpha+\beta\right)\\
& +\frac{1}{3}\frac{M_\pi^2}{M_{\eta^\prime}^2}\frac{m_s}{m}\left[\frac{m}{m_s}\sin\left(\frac{\theta}{3}-\alpha\right)+\frac{m}{m_s}\sin\left(\frac{\theta}{3}-\beta\right)+\sin\left(\frac{\theta}{3}+\alpha+\beta\right)\right]^2\Biggl.\Biggr\}\,.
\end{align*}
For $\partial V/\partial\alpha=0$, the condition for stationary points is given by
\begin{align}
0 = & \frac{m}{m_s}\sin\left(\frac{\theta}{3}-\alpha\right) - \sin\left(\frac{\theta}{3}+\alpha+\beta\right)\nonumber\\
 & -\frac{2}{3}\frac{M_\pi^2}{M_{\eta^\prime}^2}\frac{m_s}{m}\left[\frac{m}{m_s}\sin\left(\frac{\theta}{3}-\alpha\right)+\frac{m}{m_s}\sin\left(\frac{\theta}{3}-\beta\right)+\sin\left(\frac{\theta}{3}+\alpha+\beta\right)\right]\label{eq:unsolvableequation}\\
 & \qquad\qquad\ \times \left[\frac{m}{m_s}\cos\left(\frac{\theta}{3}-\alpha\right) - \cos\left(\frac{\theta}{3}+\alpha+\beta\right) \right],\nonumber
\end{align}
and accordingly for $\partial V/\partial\beta$ by replacing $\alpha\leftrightarrow\beta$, which can hardly be solved analytically with general validity.\footnote{In fact, it can, but a general expression for $\beta$ alone fills already a whole page and such an expression is barely useful in order to derive the general properties of the stationary points.}

One strategy would hence be to neglect certain terms. For that, note that ${m}/{m_s}$ and ${M_\pi^2}{/M_{\eta^\prime}^2}$ are roughly of the same order $\mathcal{O}(10^{-2})$ (using the quark and meson masses in Ref.~\cite{Tanabashi:2018oca}), so that the factor $(M_\pi^2/M_{\eta^\prime}^2)({m_s}/{m})$ is approximately $\mathcal{O}(1)$. Consequently, the eight terms in eq.~\eqref{eq:unsolvableequation} can be classified as
\begin{itemize}
\item two terms of $\mathcal{O}(1)$,
\item four terms of $\mathcal{O}(10^{-2})$, either $\sim\frac{m}{m_s}$ or $\sim\frac{M_\pi^2}{M_{\eta^\prime}^2}$, and
\item two terms of $\mathcal{O}(10^{-4})$.
\end{itemize}
The latter two can certainly be neglected but considering all of the remaining six terms gives still a barely solvable system of equations (if one considers $\partial V/\partial\beta$, too). Considering on the other hand only the two leading order terms results in the following equation
\begin{equation}\label{eq:higlyapprsolu}
\sin\left(\frac{\theta}{3}+\alpha+\beta\right)\left[\cos\left(\frac{\theta}{3}+\alpha+\beta\right)+\frac{3M_{\eta^\prime}^2m}{2M_\pi^2m_s}\right]=0\, ,
\end{equation}
which has the solutions (the expression within the square brackets can never be zero):
\begin{equation*}
\beta_{-}  =-\theta/3-\alpha\,,~~
\beta_{+}  =\beta_{-}+\pi\, .
\end{equation*}
The stationary points of the corresponding $V_{\pm}(\alpha,\theta)$ then appear at the very same $\alpha_\mathrm{I/II}^{\pm}$ as has been found for the leading order case:
\begin{equation*}
\alpha_\mathrm{I}^{+} = \frac{3\pi}{2}-\frac{\theta}{6}\,,~~
\alpha_\mathrm{II}^{+}  = \frac{\pi}{2}-\frac{\theta}{6}\,,~~
\alpha_\mathrm{I}^{-}  = -\frac{\theta}{6}\,,~~
\alpha_\mathrm{II}^{-}  = \pi-\frac{\theta}{6}\,,
\end{equation*}
with energy densities
\begin{align*}
e_\mathrm{I}^{+} (\theta) & = \Sigma m \left\{2\sin\frac{\theta}{2}+\frac{m_s}{m}-\frac{4}{3} \frac{M_\pi^2}{M_{\eta^\prime}^2} \cos^2\frac{\theta}{2} \right\}, \\
e_\mathrm{II}^{+} (\theta) & = \Sigma m \left\{-2\sin\frac{\theta}{2}+\frac{m_s}{m}-\frac{4}{3} \frac{M_\pi^2}{M_{\eta^\prime}^2}\cos^2\frac{\theta}{2} \right\} , \\
e_\mathrm{I}^{-} (\theta) & = \Sigma m \left\{-2\cos\frac{\theta}{2}-\frac{m_s}{m}-\frac{4}{3} \frac{M_\pi^2}{M_{\eta^\prime}^2}\sin^2\frac{\theta}{2} \right\} , \\
e_\mathrm{II}^{-} (\theta) & = \Sigma m \left\{2\cos\frac{\theta}{2}-\frac{m_s}{m}-\frac{4}{3} \frac{M_\pi^2}{M_{\eta^\prime}^2}\sin^2\frac{\theta}{2} \right\} .
\end{align*}
Qualitatively the behavior of the curves is entirely the same as for the leading order case given in fig.~\ref{fig:stationarypointsLO}---the curves are again centered around $\pm \Sigma m_s$. However, the appearances of the saddle points and extrema are slightly different (tab.\,\ref{tab:stationary2}). In the region around $\theta=\pi$, the points $\alpha^{-}_\mathrm{I/II}$ do not represent two local minima, but stationary points, which implies that there must be two additional solutions $\alpha^{-}_\mathrm{III\pm}$ that this time represent the two local minima that must be present in this region.
\begin{table}\centering
\begin{tabular}{lll}
\hline
\\[-1em]
$\alpha_\mathrm{I}^{+}$: & $\frac{\partial^2 V}{\partial\alpha^2},\frac{\partial^2 V}{\partial\beta^2}<0\ \forall\ \theta$ & \\[2mm]
& $\Delta^2>0\ \forall\ \theta$ & $\Rightarrow$ local maximum\\[2mm] \hline 
\\[-1em]
$\alpha_\mathrm{II}^{+}$: & $\frac{\partial^2 V}{\partial\alpha^2},\frac{\partial^2 V}{\partial\beta^2}<0\ \forall\ \theta$ & \\[2mm]
& (a) $\Delta^2>0$ for $\theta\in [0,\epsilon)$ and $\theta\in (2\pi-\epsilon,2\pi]$ & $\Rightarrow$ local maximum\\[2mm]
& (b) $\Delta^2<0$ for $\theta\in (\epsilon,2\pi-\epsilon)$ & $\Rightarrow$ saddle point
\\[2mm] \hline 
\\[-1em]
$\alpha_\mathrm{I}^{-}$: & $\frac{\partial^2 V}{\partial\alpha^2},\frac{\partial^2 V}{\partial\beta^2}>0\ \forall\ \theta$ & \\[2mm]
& (a) $\Delta^2>0$ for $\theta\in [0,\pi-\epsilon)$ & $\Rightarrow$ local minimum\\[2mm]
& (b) $\Delta^2<0$ for $\theta\in (\pi-\epsilon,2\pi]$ & $\Rightarrow$ saddle point\\[2mm] \hline 
\\[-1em]
$\alpha_\mathrm{II}^{-}$ & $\frac{\partial^2 V}{\partial\alpha^2},\frac{\partial^2 V}{\partial\beta^2}>0\ \forall\ \theta$ & \\[2mm]
& (a) $\Delta^2<0$ for $\theta\in [0,\pi+\epsilon)$ & $\Rightarrow$ saddle point\\[2mm]
& (b) $\Delta^2>0$ for $\theta\in (\pi+\epsilon,2\pi]$ & $\Rightarrow$ local minimum\\[2mm] \hline 
\end{tabular}
\caption{Classification of the stationary points of $V(\alpha,\beta,\theta)$ considering the $L_7$-term.}
\label{tab:stationary2}
\end{table}

The region, in which these two additional stationary points must appear, can be estimated by setting $\Delta^2=0$ for $\alpha^{-}_\mathrm{I/II}$, which yields
\begin{equation}\label{eq:regionofcoexistingL7}
\epsilon = \frac{8}{3}\frac{M_\pi^2}{M_{\eta^\prime}^2} \approx 0.056\, .
\end{equation}
Regarding the order of magnitude, this result is indeed comparable to the $N_f=2$ result found by Smilga (see eq.\,\eqref{eq:regionofcoexistingminimanfgleich2} above).

However, it is evident that the approximation used for this qualitative investigation does not take the effects of the mass splitting of the strange quark and the other two light quarks into account, which was the main source of non-trivial effects in the leading order case. As stated above, calculating the stationary points using eq.~\eqref{eq:unsolvableequation} and neglecting only the two most suppressed terms is still a hardly solvable task, so one may instead try to interpolate the combined effects of the $m/m_s$ mass splitting and the $L_7$-term by considering the leading order solution for $\beta_\pm$, which also approximately solves eq.~\eqref{eq:unsolvableequation}. Again, the solutions for $\alpha$ remain unchanged, but the corresponding energy densities are now given by
\begin{align*}
e_\mathrm{I}^{+} (\theta) & = \Sigma m \biggl\{\biggr.\sin\frac{\theta}{2}+\sin\left(\frac{\theta}{2}-\arcsin\left[\frac{m}{m_s}\cos\frac{\theta}{2}\right]\right)+\frac{m_s}{m}\\
& \qquad\qquad\qquad -\frac{1}{3} \frac{M_\pi^2}{M_{\eta^\prime}^2} \cos^2\left(\frac{\theta}{2}-\arcsin\left[\frac{m}{m_s}\cos\frac{\theta}{2}\right]\right)\biggl.\biggr\} \,,\\
e_\mathrm{II}^{+} (\theta) & = \Sigma m \biggl\{\biggr.-\sin\frac{\theta}{2}-\sin\left(\frac{\theta}{2}+\arcsin\left[\frac{m}{m_s}\cos\frac{\theta}{2}\right]\right)+\frac{m_s}{m}\\
& \qquad\qquad\qquad -\frac{1}{3} \frac{M_\pi^2}{M_{\eta^\prime}^2} \cos^2\left(\frac{\theta}{2}+\arcsin\left[\frac{m}{m_s}\cos\frac{\theta}{2}\right]\right)\biggl.\biggr\}\,,\\
e_\mathrm{I}^{-} (\theta) & = \Sigma m \biggl\{\biggr.-\cos\frac{\theta}{2}-\cos\left(\frac{\theta}{2}-\arcsin\left[\frac{m}{m_s}\sin\frac{\theta}{2}\right]\right)-\frac{m_s}{m}\\
& \qquad\qquad\qquad -\frac{1}{3} \frac{M_\pi^2}{M_{\eta^\prime}^2} \sin^2\left(\frac{\theta}{2}-\arcsin\left[\frac{m}{m_s}\sin\frac{\theta}{2}\right]\right)\biggl.\biggr\}\,,
\end{align*}
\begin{align*}
e_\mathrm{II}^{-} (\theta) & = \Sigma m \biggl\{\biggr.\cos\frac{\theta}{2}+\cos\left(\frac{\theta}{2}+\arcsin\left[\frac{m}{m_s}\sin\frac{\theta}{2}\right]\right)-\frac{m_s}{m}\\
& \qquad\qquad\qquad -\frac{1}{3} \frac{M_\pi^2}{M_{\eta^\prime}^2} \sin^2\left(\frac{\theta}{2}+\arcsin\left[\frac{m}{m_s}\sin\frac{\theta}{2}\right]\right)\biggl.\biggr\}
 \, ,
\end{align*}
which give once more the same shape for the corresponding curves as in fig.~\ref{fig:stationarypointsLO}. The classification of the stationary points is the same as given in table \ref{tab:stationary2}, but this time with
\begin{equation}\label{eq:regionofcoexistinginterpolated}
\epsilon = \left|-4\frac{M_\pi^2}{M_{\eta^\prime}^2}+\frac{m}{m_s}\right| \approx 0.048\, .
\end{equation}
For $m_s\to\infty$, the expression \eqref{eq:regionofcoexistinginterpolated} becomes an expression similar to \eqref{eq:regionofcoexistingL7}, where the different prefactor is certainly a consequence of the fact that both the solution to the highly approximated equation \eqref{eq:higlyapprsolu} and the interpolated solution are not exact solutions. Neglecting on the other hand the first term in \eqref{eq:regionofcoexistinginterpolated} yields as expected the leading order result $\epsilon=m/m_s$.

This brief, rather qualitative analysis clearly indicates that the effects of the $L_7$-term of the $N_f=2+1$ flavor case are similar to the effects of the corresponding $l_7$-term in the $N_f=2$ case when taking the limit $m_s\to\infty$.

\section{Summary}

Two issues related to the $\theta$-vacuum angle of QCD have been performed within this paper in order to complement and reexamine previous findings on
\begin{enumerate}
\item the vacuum structure in the large-$N_c$ limit, in particular the large-$N_c$ scaling of the cumulants of the distribution of the winding number at $\theta=0$, and,
\item the physics at $\theta\sim\pi$, in particular in the case of $2+1$ flavors.
\end{enumerate}
Section~\ref{ch:vacstruct} addressed the first topic. The case under consideration was the SU($N_f$) symmetric case for an arbitrary $N_f$ for large space-time volumes $V \Sigma m\gg 1$ within the approximation of small $\theta$. The leading and next-to-leading order expressions for the topological susceptibility $\chi_\mathrm{top}$ and the fourth cumulant $c_4$ have been derived applying both the  $\delta$-expansion and the full next-to-leading order Lagrangian in CHPT. The leading order results are in accordance with the previous findings in Refs. \cite{leutwylersmilga,lucianomeggiolaro}, the subleading  expressions represent new results.
The large-$N_c$ scaling behavior of these quantities have been worked out as
\begin{align*}
\chi_\mathrm{top} & =\mathcal{O}(1)+\mathcal{O}(N_c^{-1})\,,\\
c_4 & =\mathcal{O}(N_c^{-3})+\mathcal{O}(N_c^{-4})\, .
\end{align*}
The result for $\chi_\mathrm{top}$ has been indeed predicted in previous papers (e.\,g. \cite{kaiserleutwyler,shore2007}), and, as has been argued above, the large-$N_c$ scalings of both quantities are in accordance with what is allowed from general large-$N_c$ counting rules for $e_\mathrm{vac}(\vartheta)$ (cf.\,eq.\,\eqref{eq:evacexpansionlargeNc}). The conclusion is thus that in the large-$N_c$ limit the distribution of the winding number at $\theta=0$ becomes  Gaussian, since the fourth cumulant $c_4$ is strongly suppressed as well as higher cumulants.

The findings seem, however, to be in conflict with what has been reported in Refs. \cite{vicari,bonati} with respect to lattice simulations. The values for $\chi_\mathrm{top}$ and $c_4$ measured on the lattice seem to corroborate their assumed scaling behavior $\chi_\mathrm{top}=\mathcal{O}(1)+\mathcal{O}(N_c^{-2})$ and $c_4=\mathcal{O}(N_c^{-2})+\mathcal{O}(N_c^{-4})$, but there are several issues related to these claims: (a) The authors refer to the same large-$N_c$ counting rules and the same equation for $e_\mathrm{vac}(\vartheta)$, 
however, as has been argued in the corresponding section, this formula gives only an upper bound for the large-$N_c$ scalings of $\chi_\mathrm{top}$ and $c_4$. (b) The amount of measurements is still small, while the errors are sizable, so one cannot conclusively deduce the exact scaling of $\chi_\mathrm{top}$ and $c_4$ from the data available so far. (c) The best fits to the lattice data are achieved by those derived here  for $\chi_\mathrm{top}$ and $c_4$ in the $\delta$-expansion. The results derived in the present study are hence very well also consistent with the recent results from lattice simulations, though it is of course desirable that future research provides more (and more precise) lattice data in order to confirm our finding. 

Taking up a prior work conducted by Smilga~\cite{smilga} in which the author explores the physics at $\theta$ around $\pi$ in the case of two and three mass-degenerate flavors, the second part of this paper (section\,\ref{ch:thetasimpi}) extended this previous findings considering two light flavors with a degenerate mass $m$ as well as the heavier strange quark with a mass $m_s\gg m$. Smilga demonstrated that the region $[\theta-\epsilon,\theta+\epsilon]$ of two coexisting minima in the case of two light flavors of equal mass is very small, roughly $\epsilon=2M_\pi^2/(3M_\eta^2)\approx 0.04$, while it is very extended in the case of three mass-degenerate flavors, where $\epsilon=\pi/2$. 

The results for the $m\ll m_s$ approximation clearly reconfirm these findings. First, it has been shown that within the region of two coexisting minima two additional stationary points show up, which can of course be expected: the ``valley'' between the high laying regions of the potential's landscape for $\alpha,\beta\in [0,2\pi]$ consists of one saddle point as long as there is only one absolute minimum. Once a second local minimum appears, the ``valley'' must consist of two saddle points, too. Interestingly, in the leading order case, where the stationary points $\alpha_{\mathrm{I/II}}^{-}$ represent the two coexisting local minima, the saddle points are located at the two additional stationary points $\alpha_{\mathrm{III}\pm}^{-}$, whereas in the case including the $L_7$-term the points $\alpha_{\mathrm{I/II}}^{-}$ turn to saddle points in the region $[\theta-\epsilon,\theta+\epsilon]$, so that the additional solutions $\alpha_{\mathrm{III}\pm}^{-}$ must represent local minima.

The region of two local minima at leading order is $\epsilon= \arcsin({m}/{m_s})$, hence tending to $\epsilon\to{\pi}/{2}$ for $m_s\to m$, whereas $\epsilon\to 0$ for $m_s\to \infty$. The latter corresponds to the insight that the leading order potential in the SU(2) symmetric case does not depend on the matrix $U$, so in order to bring the $2+1$ flavor case into contact with the results for two mass-degenerate flavors one has to consider contributions from the next-to-leading order effective Lagrangian, too. While we have not given exact solutions for the potential including the $L_7$-term, the qualitative analysis indicated that the order of magnitude of the contributions stemming from this term is comparable ($\epsilon={8}{M_\pi^2}/(3{M_{\eta^\prime}^2})\approx 0.056$) to those that stem from the corresponding $l_7$-term in the $N_f=2$ case. The combined effects of the leading order potential respecting the mass splitting $m/m_s$ and the $L_7$-term lead to a somewhat smaller region $\epsilon = |-4{M_\pi^2}/{M_{\eta^\prime}^2}+{m}/{m_s}| \approx 0.048$.

In addition, the profile of the energy barrier appearing between the domains of the two degenerate vacuum states at $\theta=\pi$ has been calculated to leading order. The domain wall tension is given by
\[
T_W= 2F^2 M_\pi \sqrt{\frac{m}{m_s}}\,.
\]
From that, the decay rate of the metastable vacuum has been estimated showing likewise a good agreement with Smilga's findings. The additional factor $(m/m_s)^2$ appearing in the exponential function leads to a shorter lifetime than that for the $N_f=3$ mass-degenerate case, the overall lifetime is, however, still large mainly because of the factor $1/|\theta^\prime|^3$ in the exponential, where $0<|\theta^\prime|\ll 1$.

\acknowledgments

We thank Leonardo Giusti for some useful comments.
This work is supported in part by the National Natural Science Foundation of China (NSFC) and  the Deutsche Forschungsgemeinschaft (DFG) through the funds provided to the Sino-German Collaborative Research Center ``Symmetries and the Emergence of Structure in QCD"  (NSFC Grant No. 11621131001, DFG Grant No. TRR110), by the NSFC under Grant No. 11747601 and No. 11835015, by the Chinese Academy of Sciences (CAS) under Grant No. QYZDB-SSW-SYS013 and No. XDPB09, by
the CAS Center for Excellence in Particle Physics (CCEPP),  by the CAS President's International Fellowship Initiative (PIFI) (Grant No.~2018DM0034), and by the VolkswagenStiftung (Grant No. 93562).


\providecommand{\href}[2]{#2}\begingroup\raggedright\endgroup

\end{document}